\documentclass[manuscript]{aastex}

\usepackage{comment}
\usepackage{url}

\usepackage{color}
\definecolor{def}{rgb}{0,0,0}
\definecolor{del}{rgb}{0.5, 0.5, 1}
\definecolor{del2}{rgb}{0.5, 0.5, 1}
\definecolor{RC1}{rgb}{0,0,0}
\definecolor{RC2}{rgb}{0,0,0}
\definecolor{RC3}{rgb}{0,0,0}
\definecolor{RC4}{rgb}{0,0,0}
\definecolor{RC5}{rgb}{1,0,0}

 \newcommand{\CO}{${\rm ^{12}CO}$}
 \newcommand{\tCO}{${\rm ^{13}CO}$}
 \newcommand{\CeO}{${\rm C^{18}O}$}
 \newcommand{\Rtt}{$R^{13/12}_{2-1}$}
 \newcommand{\Rul}{$R^{13}_{2-1/1-0}$}
 \newcommand{\CI}{\mbox{\ion{C}{1}}}
 \newcommand{\HII}{\mbox{\ion{H}{2}}}
 \newcommand{\Ha}{H$\alpha$}

\shorttitle{Physical properties of Orion molecular cloud}
\shortauthors{Nishimura et al.}

\begin{document}


\title{Revealing the physical properties of molecular gas in Orion with a large scale survey in $J$ = 2--1 lines of $^{12}$CO, $^{13}$CO and C$^{18}$O}


\author{
Atsushi Nishimura\altaffilmark{1},
Kazuki Tokuda\altaffilmark{1},
Kimihiro Kimura\altaffilmark{1},
Kazuyuki Muraoka\altaffilmark{1},
Hiroyuki Maezawa\altaffilmark{1},
Hideo Ogawa\altaffilmark{1},
Kazuhito Dobashi\altaffilmark{2},
Tomomi Shimoikura\altaffilmark{2},
Akira Mizuno\altaffilmark{3},
Yasuo Fukui\altaffilmark{4},
}
\and
\author{Toshikazu Onishi\altaffilmark{1}}

\altaffiltext{1}{Department of Physical Science, Graduate School of Science, Osaka Prefecture University, 1-1 Gakuen-cho, Naka-ku, Sakai, Osaka 599-8531, Japan}
\email{s\_a.nishimura@p.s.osakafu-u.ac.jp}

\altaffiltext{2}{Department of Astronomy and Earth Sciences, Tokyo Gakugei University, 4-1-1 Nukuikita-machi, Koganei, Tokyo 184-8501, Japan}
\altaffiltext{3}{Solar-terrestrial Environment Laboratory, Nagoya University, Furo-cho, Chikusa-ku, Nagoya, Aichi 464-8601, Japan}
\altaffiltext{4}{Department of Physics and Astrophysics, Nagoya University, Furo-cho, Chikusa-ku, Nagoya, Aichi 464-8602, Japan}


\begin{abstract}
We present fully sampled $\sim3 \arcmin$ resolution images of the \CO ($J$ = 2--1), \tCO ($J$ = 2--1), and \CeO ($J$ = 2--1) emission taken with the newly developed 1.85-m mm-submm telescope toward the entire area of the Orion A and B giant molecular clouds.
The data were compared with the $J$ = 1--0 of the \CO, \tCO, and \CeO \ data taken with the Nagoya 4-m telescope and the NANTEN telescope at the same angular resolution to derive the spatial distributions of the physical properties of the molecular gas.
We explore the large velocity gradient formalism to determine the gas density and temperature by using the line combinations of \CO($J$ = 2--1), \tCO($J$ = 2--1), and \tCO($J$ = 1--0) assuming uniform velocity gradient and abundance ratio of CO.
The derived gas density is in the range of 500 to 5000 cm$^{-3}$, and
the derived gas temperature is mostly in the range of 20 to 50 K along the cloud ridge with a temperature gradient depending on the distance from the star forming region.
We found the high-temperature region at the cloud edge facing to the \HII \ region, indicating that the molecular gas is interacting with the stellar wind and radiation from the massive stars.
In addition, we compared the derived gas properties with the Young Stellar Objects distribution obtained with the Spitzer telescope to investigate the relationship between the gas properties and the star formation activity therein.
We found that the gas density and star formation efficiency are well positively correlated, indicating that stars form effectively in the dense gas region.
\end{abstract}





\section{Introduction}

Most of stars are formed in Giant Molecular Clouds (GMCs) in the Galaxy (e.g., \citealt{lad98}).
Molecular rotational transitions have been used to investigate the physical properties of the molecular gas to be compared with the star formation activities therein. 
The lowest transition ($J$ = 1--0) lines of CO and its isotopes are good mass (or column density) tracers of molecular gas from relatively low-density regions ($\sim10^{2}$ cm$^{-3}$) to high-density regions ($\ga 10^{4}$ cm$^{-3}$). 
This is because the molecule is the most abundant in the interstellar medium (ISM) except for molecular hydrogen and helium, and also because the Einstein coefficient for spontaneous emission, $A_{10}$, is small,
for which CO can be excited by collision even at relatively low density, making the low rotational transitions of CO good probes of the molecular gas.
The $J$ = 1--0 lines of \CO \ and \tCO \ have, therefore, been used to carry out large-scale observations covering the large areas of various nearby star forming sites (e.g., \citealt{dam01, dob94, tac96, miz98, kaw98, oni99, yon05, rid06, jac06, gol08}).
Such large-scale surveys have provided us invaluable information to characterize the morphological and physical properties of molecular clouds. 
Meanwhile, the higher excitation lines such as CO($J$ = 2--1) have been used to determine the local densities and temperatures by making use of the fact that they have higher critical densities for the excitation, which are also quite important for us to diagnose the evolutionary status of the molecular clouds (e.g., \citealt{sak95, beu00, yod10, pol12}).
However, such observations were conducted only at coarse angular resolutions or only toward small regions, mainly because the development of sensitive receivers at the high frequencies had been very difficult and because the opacity of the earth atmosphere is high at low altitude sites.

The Orion star forming region contains nearest GMCs with massive star clusters, and then it is one of the most suitable site to investigate the process of star formation and the effect on the parent cloud.
It includes two GMCs, Orion A and Orion B, whose distance is estimated to be 410 pc (e.g., \citealt{men07, san07, hir07}).
Extensive observations of the whole Orion region have been made in
\CO ($J$ = 1--0) \citep{kut77, mad86, wil05},
\CO \ and \tCO ($J$ = 1--0) \citep{rip13},
\CO ($J$ = 2--1) \citep{sak94},
\CO ($J$ = 3--2) and \CI($^3P_1$--$^3P_0$) \citep{ike02}.
Those of the Orion A cloud have been made in
\CO ($J$ = 1--0) \citep{shi11, nak12},
\tCO ($J$ = 1--0) \citep{bal87, nag98},
\tCO \ and \CeO ($J$ = 3--2) \citep{buc12},
CS($J$ = 1--0) \citep{tat93},
CS ($J$ = 2--1) \citep{tat98},
and  H$^{13}$CO$^+$($J$ = 1--0) \citep{ike07}.
Those of the Orion B cloud have been made in 
\CeO ($J$=1--0) and H$^{13}$CO$^+$($J$ = 1--0) \citep{aoy01},
\tCO \ and \CeO ($J$ = 3--2) \citep{buc10},
CS ($J$ = 2--1) \citep{lad91},
and  H$^{13}$CO$^+$($J$ = 1--0) \citep{ike09}.
These observations revealed that the clouds are full of filaments and cores \citep{nag98, aoy01} and are affected by UV radiations from the nearby OB stars \citep{bal87, wil05}. 
The northern part of the Orion A cloud and the entire Orion B cloud are exposed to the strong UV radiation field of $G_0 = 10^{4-5}$ \citep{tie85, kra96}.
On the other hand, the central and southern parts of the Orion A are of low UV field and show quiescent low-mass star formation.
The difference in the activity of the star formation should result in different physical properties of the molecular gas.

\citet{sak94} carried out a large area \CO($J$ = 2--1) mapping of the Orion A and Orion B clouds, and compared them with the \CO($J$ = 1--0) data obtained by \citet{mad86} on the same observing grids at a same angular resolution of 9\arcmin.
They observed a systematic variation of the \CO($J$ = 2--1)/\CO($J$ = 1--0) intensity ratio over the entire extents of the GMCs, reflecting the physical properties of the molecular gas there. 
It was, however, difficult to derive the properties precisely, especially toward the ridge area where star formation is taking place because the optical depth toward the ridge is expected to be very large for the \CO \ emission. 
The angular resolution (9\arcmin) corresponds to a spatial resolution of $\sim 1$ pc at the distance of the Orion clouds. 
Because the Jeans length of the gas with $n(\rm H_2) \sim$ a few $\times$ 100 cm$^{-3}$ and $T \sim 10$ K is estimated to be $\sim1$ pc, observations with a spatial resolution of $<$ 1 pc are needed to investigate the physical properties of the individual clouds and the dynamical state.  

We developed a 1.85-m mm-submm telescope for large-scale molecular observations in $J$ = 2--1 lines of \CO, \tCO, and \CeO \ \citep{oni13}. 
The purpose of the telescope is to reveal the physical properties of the molecular clouds extensively at an angular resolution of $\sim$3\arcmin. 
As one of the major survey projects of the telescope, we have carried out a full-sampling observation of both the Orion A and Orion B clouds, and compared them with the data in $J$ = 1--0 lines taken by the 4-m telescopes of Nagoya University.  
This paper is organized as follows: in Section 2, the observations and data reduction procedures of the 1.85-m telescope and the 4-m telescopes are described.
In Section 3, we present results of CO($J$ = 2--1) and CO($J$ = 1--0) observations.
In Section 4, we describe our analyses and present the derived physical properties of the Orion molecular clouds.
In Section 5, we discuss the cloud properties, star formation activity of this region, and the surrounding environment.
Finally we summarize the paper in Section 7.



\section{Observations}
\label{sec_observation}


\subsection{\CO($J$ = 2--1), \tCO($J$ = 2--1), and \CeO($J$ = 2--1)}

Observations of the $J$=2--1 transitions of $^{12}$CO, $^{13}$CO, and C$^{18}$O were carried out with the 1.85-m telescope installed at Nobeyama Radio Observatory (NRO) which is enclosed in a radome that prevents the telescope structure distortion due to outdoor conditions (e.g., precipitation, wind, and sunlight).
At 230 GHz, the telescope has a beam size of 2\farcm 7 (HPBW)
which was measured by continuum scans of the Jupiter.
We used a two sideband separating (2SB) superconductor-insulator-superconductor (SIS) mixer receiver to observe $J$ = 2--1 lines of $^{12}$CO, $^{13}$CO, and C$^{18}$O simultaneously.
The typical noise temperature of the receiver  $T_{\rm RX}$ was measured to be $\sim100$ K (single sideband) and the image rejection ratio (IRR) was measured to be 10 dB or higher.
A Fast Fourier Transform (FFT) spectrometer with 1 GHz bandwidth and 61 kHz frequency resolution is installed as the backend system.
We used the spectrometer for the observations in the three lines by dividing the frequency band into three parts.
Each part has a velocity coverage and a velocity resolution of $\sim$250 km s$^{-1}$ and 0.08 km s$^{-1}$, respectively.
Further information of the telescope is described by \citet{oni13}.

The intensity calibration was carried out by observing a standard source, Orion KL, as described in \citet{oni13}.
They estimated the uncertainty of the calibration to be $\sim$10\%.
The other factor that may affect the calibration accuracy is the beam coupling to the sources with different extents.
Figure 4 of \citet{oni13} shows that there is no large-scale deformations affecting the strength of the error beam, and the main dish was made by monobloc casting, which has no effect of small-scale fluctuations of the surface like misalignments of panels sometimes seen in large telescopes (e.g., \citealt{gre98} for the case of IRAM 30m).
\citet{oni13} also showed that the beam is nearly circular symmetry without significant minor lobes observed.
The typical antenna temperature toward the Orion KL is $\sim$45 K in \CO($J$=2--1) after the correction of the effect of the spillover to the image sideband.
The brightness temperature of the Orion KL is 63 K in \CO($J$=2--1) \citep{oni13}.
Therefore, the typical scaling factor from the antenna temperature to $T_{\rm R}^{*}$ is 1/0.7.
The moon efficiency was measured to be $\sim$70\% with an error of $\sim$10\%. 
All of these facts indicate that the calibration error due to the beam coupling to the sources with different extents is smaller than that for the intensity calibration to the $T_{\rm R}^{*}$ scale, which is $\sim$10\% \citep{oni13}.
Therefore, the uncertainty of in the calibration is estimated to be $\sim$10\% here.

The observations were carried out from 2011 January to 2011 May.
The \CO, \tCO, and \CeO \  lines were observed simultaneously.
The system noise temperatures including the atmospheric attenuation $T_{\rm sys}$ were in the range of 200 to 400 K for the three lines.
We have covered 55 deg$^{2}$ around the Orion A and Orion B molecular clouds. 
The area was divided into 55 submaps of $1^{\circ} \times 1^{\circ}$.
We observed each submap using the on-the-fly (OTF) mapping technique along the galactic coordinates.
The scan data were obtained with a fully sampled grid of 1\arcmin.
We selected 30 different OFF positions toward where we confirmed that the \CO \ emission is absent at the rms noise level of $\sim 0.1$ K at a velocity resolution of 0.08 km s$^{-1}$.
In this paper, we use the calibrated $T^*_{\rm R}$ scale \citep{kut81}.
Before observing each submap, we observed the Orion KL ($\alpha_{\rm J2000} = 05^{\rm h}35^{\rm m}14 \fs 46, \delta_{\rm J2000} = -05 \arcdeg 22 \arcmin 29 \farcs 6$) for an intensity calibrations to $T^*_{\rm R}$ scale by assuming its peak temperature of \CO($J$ = 2--1) is 63 K \citep{oni13}.
We applied each scale factor obtained by the \CO \ observations for the intensity calibrations of \tCO \ and \CeO.
We subtracted a polynomial curve from each spectrum to ensure the linear baseline, and resampled the raw OTF data  onto the 1' grid by convolving them with a Gaussian function.
The rms noise of the resulting data, $\Delta T^*_{\rm R}$, is typically $\sim0.45$ K at the velocity resolution of 0.3 km s$^{-1}$ with an effective beam size of 3\farcm 4.
In addition, we made a moment masked cube (e.g., \citealt{dam11}) to suppress the noise effect in the velocity analyses.
The moment masked cube has zero values at the emission free pixels, which is useful to avoid a large error arising from the random noise.
The emission free pixels are determined by identifying significant emission from the smoothed data whose noise level is much lower than the original data.

%
%


\subsection{\CO($J$ = 1--0), \tCO($J$ = 1--0), and \CeO($J$ = 1--0)}

The \CO ($J$=1--0) and \tCO ($J$=1--0) data were taken with the two 4-m millimeter-wave telescopes at Nagoya University \citep{kaw85, fuk91}.
The beam size of the telescopes were 2\farcm 7 (HPBW) at 110 GHz.
Each telescope was equipped with a 4 K cooled superconducting mixer receiver \citep{oga90}, which provided typical single sideband system noise temperatures of $\sim$400 K and $\sim$150 K for \CO \ and \tCO \ frequency bands, respectively, including the atmospheric attenuation.
The spectrometers were acousto-optical spectrometers (AOSs) with the 40 MHz bandwidth and 40 kHz frequency resolution, corresponding to a velocity coverage and resolution of 100 and 0.1 km s$^{-1}$, respectively.
The data were obtained by frequency switching mode with a grid spacing of 4\arcmin \ and 2\arcmin.
The rms noise level is better than 0.5 K in $T^*_{\rm R}$ scale.
The survey data were partially published by \citet{nag98} for \tCO ($J$=1--0) data of the Orion A.

The \CeO ($J$=1--0) data were taken with the NANTEN 4-m telescope \citep{miz04} which is equipped with the same receiver and spectrometer as the Nagoya University 4-m telescopes described above.
The \CeO ($J$=1--0) observations were carried out toward the region where the \tCO ($J$ = 1--0) line emission is strong.
The data were obtained by frequency switching mode at a grid spacing of 2\arcmin.
The rms noise level is better than 0.1 K in $T^*_{\rm R}$ scale.
The survey data were partially published by \citet{aoy01} for the observation of the Orion B.



\section{Results}



\subsection{Spatial distribution}


\subsubsection{\CO($J$ = 2--1) and \CO ($J$ = 1--0)}

\label{sec_12CO}

\begin{figure}
\plotone{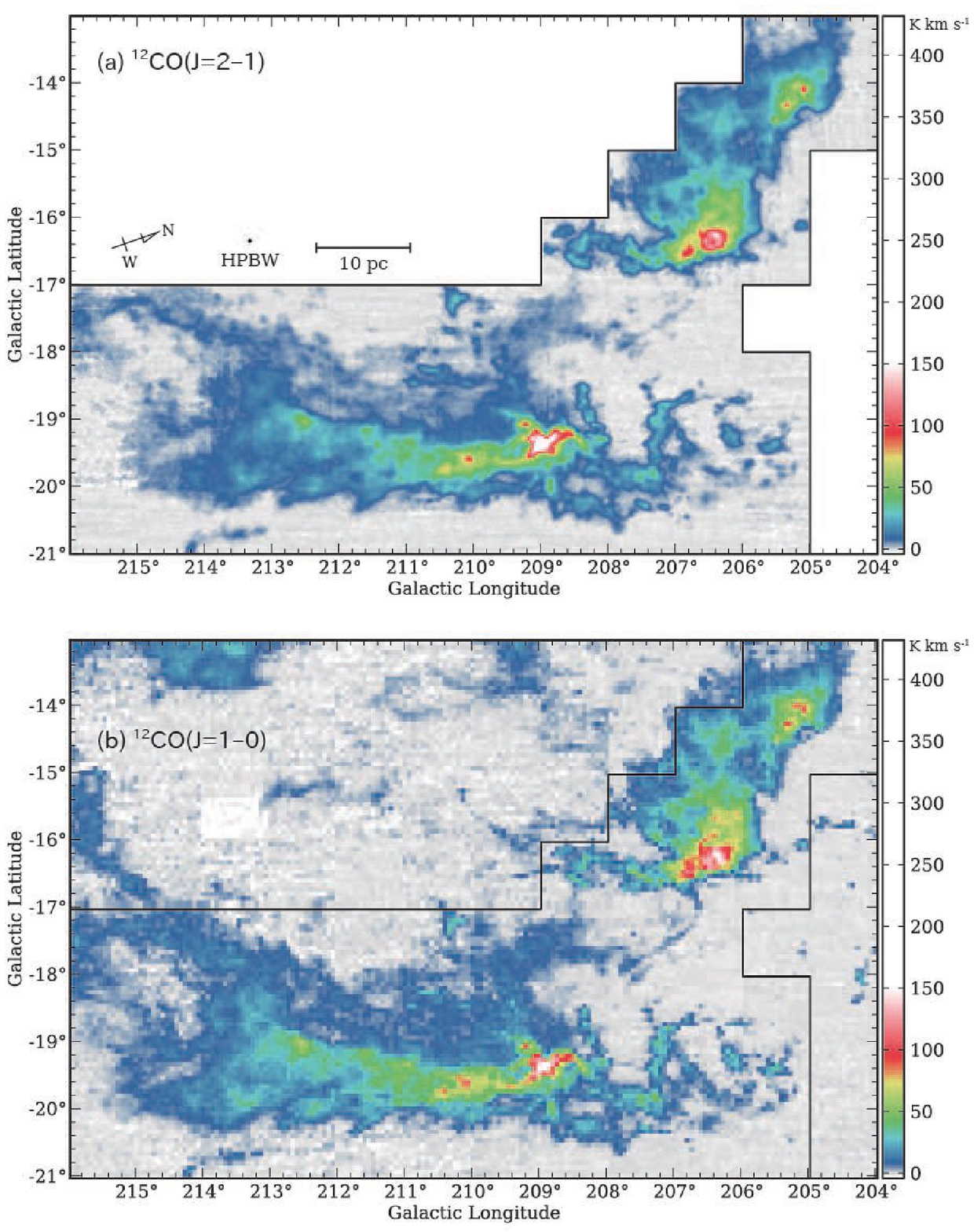}
\caption{
Integrated intensity maps of (a)\CO($J$ = 2--1) with peak intensity of 431 K km s$^{-1}$ and (b)\CO($J$ = 1--0) with peak intensity of 359 K km s$^{-1}$ toward the Orion A and B molecular clouds.
The velocity range used for the integration is $0 < V_{\rm LSR} < 20$ km s$^{-1}$ for both of the maps.
The area indicated by the solid line denotes the field observed with the 1.85-m telescope.
(A color version of this figure is available in the online journal.)
\label{fig_ii_12CO}}
\end{figure}

\begin{figure}
\plotone{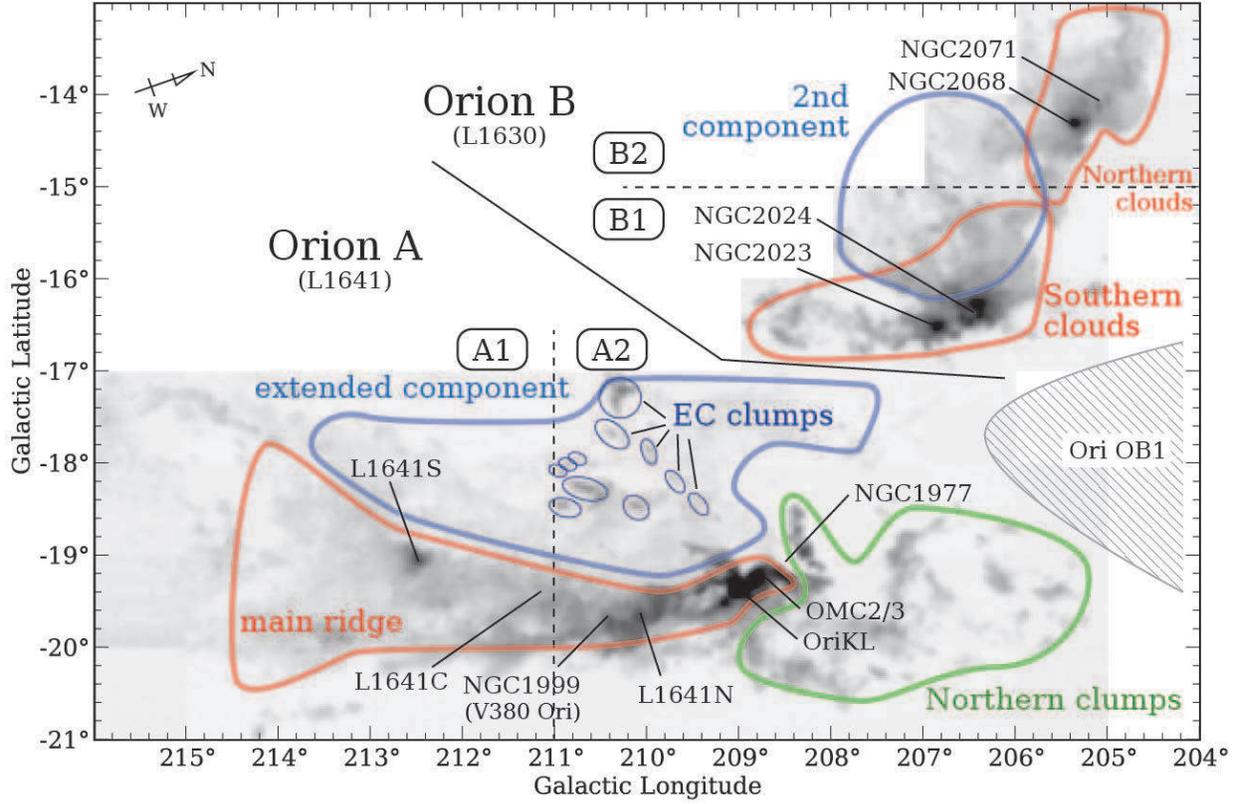}
\caption{
Explanatory map of the \CO \ emission features.
Gray scale is the peak intensity distributions of the \CO($J$ = 2--1) emission ranging from 0.5 to 25 K.
Details of the features are described in subsection \ref{sec_12CO}.
(A color version of this figure is available in the online journal.)
\label{fig_finding}}
\end{figure}

Figure \ref{fig_ii_12CO} shows velocity-integrated intensity maps of \CO($J$ = 2--1) and \CO($J$ = 1--0) observed with the 1.85-m telescope and the 4-m telescopes, respectively.
The intensities are calculated by integrating the spectra between $v_{\rm LSR}$ = 0 and 20 km s$^{-1}$ where the emission exists.
The Orion A and B molecular clouds are fully covered with significantly improved sensitivity, angular and frequency resolutions compared with previous \CO($J$ = 2--1) observations carried out by \citet{sak94}.
As pointed out by \citet{sak94}, we found that the two transitions of \CO \ exhibit a similar spatial distribution on a large scale.
However, small-scale differences
are seen in the lower intensity regions.
Actually, in the higher intensity regions ($>100$ K km s$^{-1}$) both images exhibit almost similar distributions, while in the lower intensity regions ($<10$ K km s$^{-1}$) the $J$ = 1--0 emission is apparently more widely distributed than $J$ = 2--1 emission.
In the following, we describe the spatial distribution for the Orion A and B in more detail (see Figure \ref{fig_finding}).
In the following, we briefly summarize 

The Orion A molecular cloud is distributed almost parallel to the galactic plane at $b = -19 \fdg 5$.
The maximum intensities of \CO($J$ = 2--1) and \CO($J$ = 1--0) are both found at the position of Orion KL ($l = 208 \fdg 98, b = -19 \fdg 36$) whose peak temperatures are 62.6 K and 55.5 K, respectively.
The Orion A molecular cloud can be divided into three physically different regions: the main ridge, the extended component, and northern clumps.
The main ridge is a major component of the Orion A cloud including Integral Shaped Filament (ISF; \citealt{bal87}), L1641, and other star forming sites.
The main ridge consists of a number of clumps, filaments \citep{bal87, nag98, nak12}, and shells \citep{hey92, nak12} and most of the structures are also observed in the present survey. 
One of the noticeable features in the main ridge is the existence of the gradient of some physical parameters including the line center velocity (e.g., \citealt{mad86}), volume density \citep{sak94}, excitation temperature, and filament width \citep{nag98}, which we will discusse in the following subsections (\S \ref{sec_velocity}, \S \ref{sec_ratio}).
Another striking feature is the well-defined boundary observed on the western side of the main ridge. 
This feature is observed by \citet{wil05} with a 9\arcmin \ resolution and they suggested the boundary is due to the stellar wind and/or radiation from the Orion OB1 association or ancient interaction with supernovae.
The extended component (EC; \citealt{sak97}) is located in the eastern side of the main ridge with the less intense emission typically $<15$ K km s$^{-1}$ in the integrated intensity map of \CO($J$ = 2--1). 
Molecular gas of the EC has observed only at a coarse angular resolution \citep{wil05}, or toward small regions \citep{sak97}.
\citet{sak97} proposed that the EC is located in front of the main ridge, and it was formed as a result of the interaction between the galactic atomic gas and the dense molecular gas in the main ridge.
In the present survey, we covered the entire extent of the EC at the higher angular resolution.
We detected a dozen of clumps which have relatively high intensity and well-defined boundary toward the EC region(hereafter "EC clumps").
The remarkable feature of the northern clumps are the lack of diffuse emission probably due to the interaction with the surrounding OB associations.

The Orion B molecular cloud is located in the upper-right side in the Figure \ref{fig_ii_12CO}.
The maximum intensity  of \CO($J$ = 2--1) is observed toward NGC2068 ($l = 205 \fdg 37, b = -14 \fdg 33$) with a peak temperature of 31.9 K and that of \CO($J$ = 1--0) is observed toward NGC2023 ($l = 206 \fdg 87, b = -16 \fdg 53$) with a peak temperature of 35.4 K.
The Orion B molecular cloud can be divided into three regions: the southern part including NGC2023 and NGC2024 (hereafter, we call this part Southern cloud), the northern part including NGC2068 and NGC2071 (hereafter, Northern cloud), and the central part which has only diffuse extended emission (hereafter, 2nd component).
The Southern cloud and the Northern cloud have clear boundary in the direction of the Orion OB1 association, which may be due to the stellar wind and/or radiation from massive stars.
The 2nd component has a different velocity component from the Northern and Southern clouds, and thus it seems to have no physical relation to other clouds \citep{mad86}.


\subsubsection{\tCO (J = 2--1) and \tCO (J = 1--0)}

\begin{figure}
\plotone{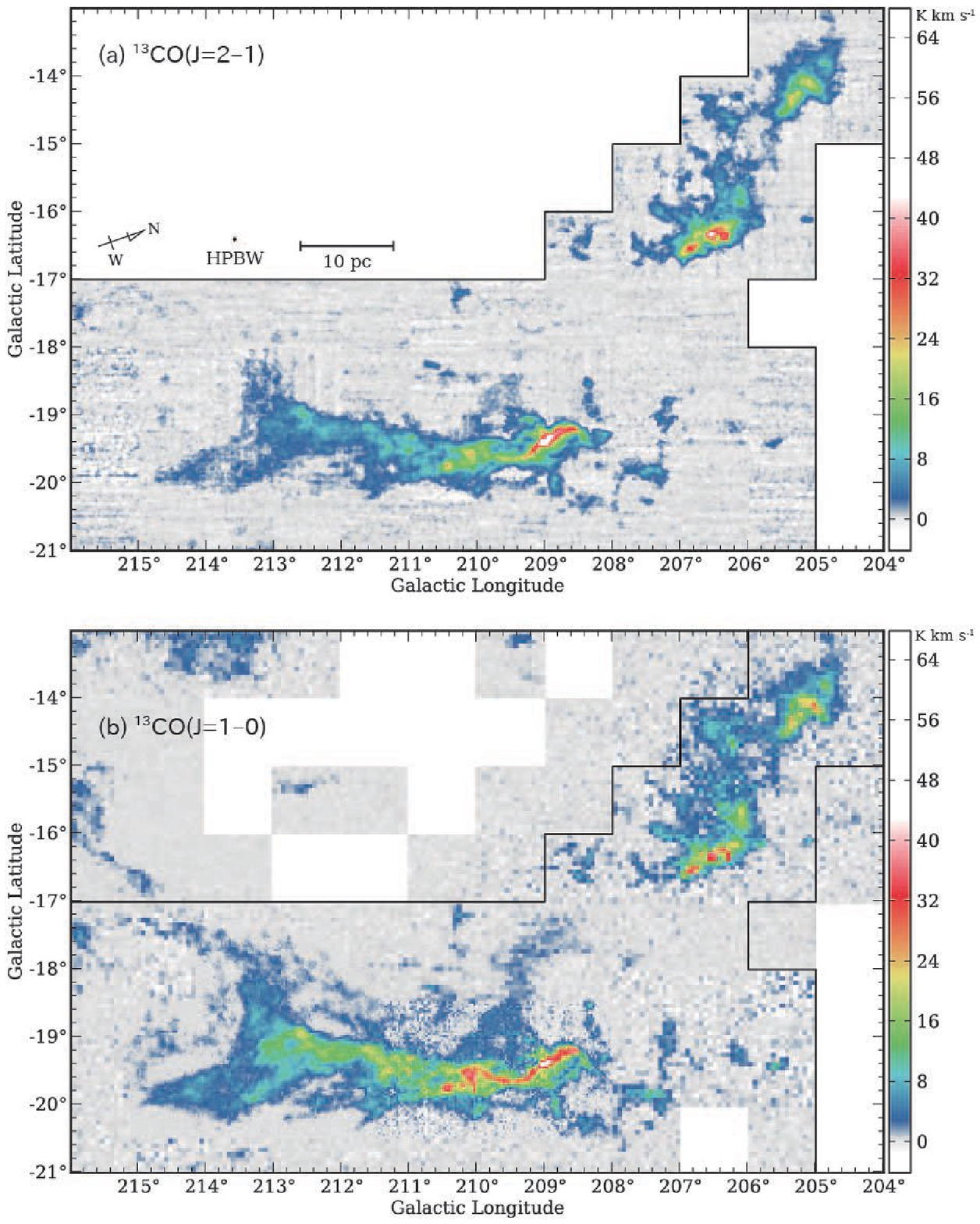}
\caption{
Integrated intensity maps of (a)\tCO($J$ = 2--1) with peak intensity of 68 K km s$^{-1}$ and (b)\tCO($J$ = 1--0) with peak intensity of 48 K km s$^{-1}$ toward the Orion A and B molecular clouds.
The velocity range used for the integration is $0 < V_{\rm LSR} < 20$ km s$^{-1}$ for both of the maps.
The area indicated by the solid line denotes the field observed with the 1.85-m telescope.
(A color version of this figure is available in the online journal.)
\label{fig_ii_13CO}}
\end{figure}

Figure \ref{fig_ii_13CO} shows velocity-integrated intensity maps of \tCO($J$ = 2--1) and \tCO($J$ = 1--0).
In general, both the $J$ = 2--1 and $J$ = 1--0 lines have similar spatial distributions except for the lower intensity region around the main ridge.
The \tCO \ emission is detected toward the region where \CO \ emission is relatively strong.
However, the \tCO($J$ = 2--1) emission is not detected in the regions toward with extended week \CO \ emission.
In the Orion A, the maximum intensity of \tCO($J$ = 2--1) is observed toward Orion KL with a peak temperature of 17.4 K and that of \tCO($J$ = 1--0) is observed toward 10\arcmin \ north to Orion KL ($l = 208 \fdg 80, b = -19 \fdg 27$) with a peak temperature of 12.8 K.
The main ridge exhibits more filamentary shape than the \CO \ distributions, which is considered to reflect the inner structure of the clouds due to its smaller optical depth.
The main ridge has almost constant intensity ($\sim10$ K km s$^{-1}$) expect for the local peaks around the L1641N ($l = 210 \fdg 1, b = -19 \fdg 6$).
The helix shaped structure is seen in the southern side of the main ridge between $l = 211 \arcdeg$ and $213 \arcdeg$, representing the possible influence of the magnetic field \citep{uchi91}.
The main ridge has well-defined boundary on both the western and eastern side of the filament.
The diffuse emission is not seen toward the EC region in the $J$ = 2--1 emission, while some of the EC clumps are clearly detected.
In the northern clumps region, \tCO \ is observed where \CO \ is relatively strong.
In the Orion B, the maximum intensity of \tCO($J$ = 2--1) is observed toward NGC2024  ($l = 206 \fdg 57, b = -16 \fdg 37$) with a peak temperature of 16.7 K and that of  \tCO($J$ = 1--0) is observed toward NGC2023 ($l = 206 \fdg 87, b = -16 \fdg 60$) with a peak temperature of 14.4 K.
In the $J$ = 2--1 emission, the Southern cloud and the Northern cloud are  clearly separated.
The clouds have well-defined boundary toward the western direction while some diffuse components are extended toward the opposite direction.


\subsubsection{\CeO (J = 2--1) and \CeO (J = 1--0)}

\begin{figure}
\plotone{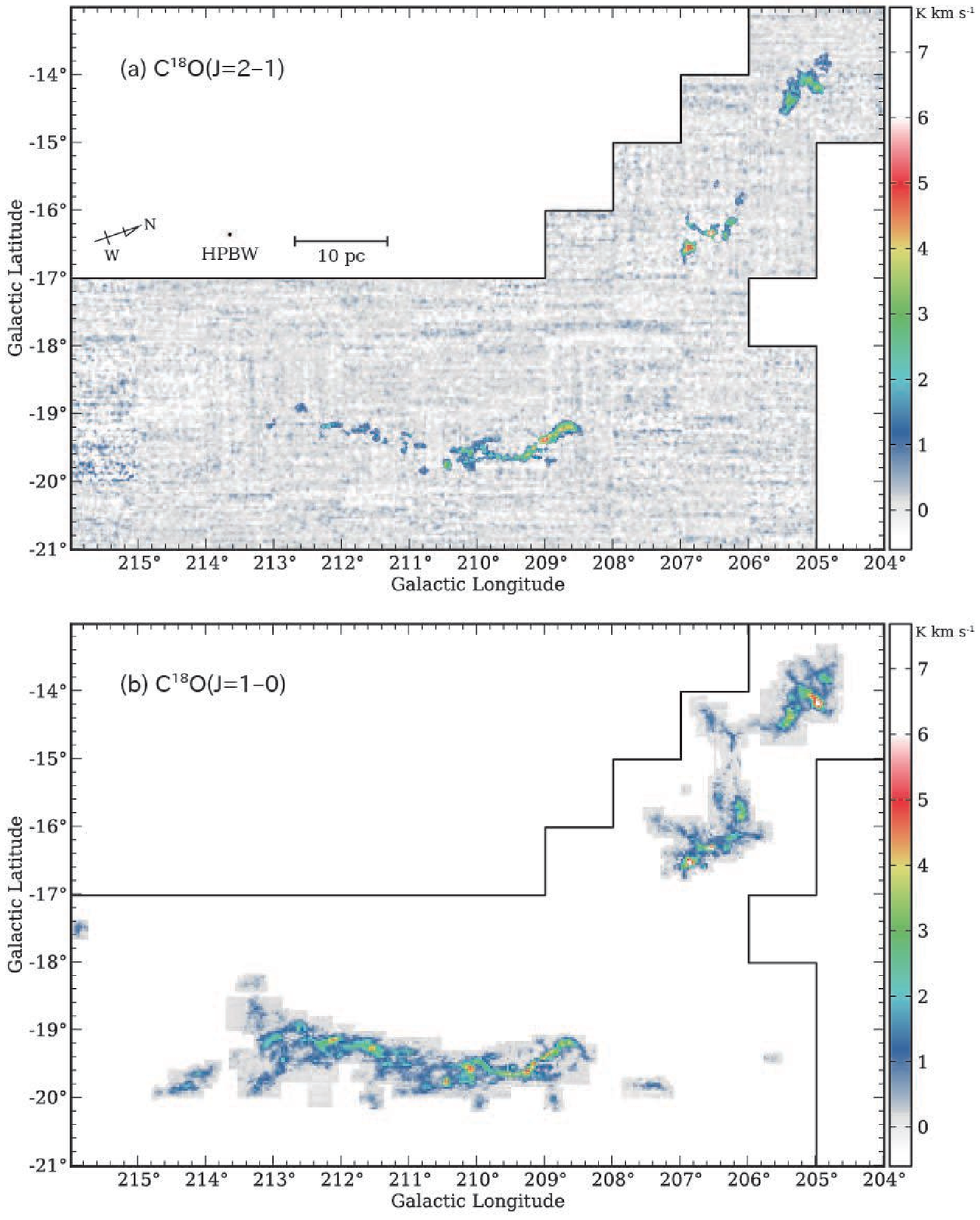}
\caption{
Integrated intensity maps of (a)\CeO($J$ = 2--1) with peak intensity of 7.7 K km s$^{-1}$ and (b)\CeO($J$ = 1--0) with peak intensity of 8.0 K km s$^{-1}$ toward the Orion A and B molecular clouds.
The velocity range used for the integration is $0 < V_{\rm LSR} < 20$ km s$^{-1}$ for both of the maps.
The area indicated by the solid line denotes the field observed with the 1.85-m telescope.
(A color version of this figure is available in the online journal.)
\label{fig_ii_C18O}}
\end{figure}

Figure \ref{fig_ii_C18O} shows velocity integrated intensity maps of \CeO($J$ = 2--1) and \CeO($J$ = 1--0).
In the Orion A, the maximum intensity of $J$ = 2--1 is observed toward 14\arcmin \ north to Orion KL ($l = 208 \fdg 78, b = -19 \fdg 23$) with a peak temperature of 3.3 K, and that of $J$ = 1--0 is observed toward L1641S ($l = 212 \fdg 10, b = -19 \fdg 17$) with a peak temperature of 2.8 K.
In the Orion B, the maximum intensities of $J$ = 2--1 and $J$ = 1--0 are observed toward NGC2023 ($l = 206 \fdg 87, b = -16 \fdg 57$) with peak temperatures of 3.8 K and 3.6 K, respectively.
The \CeO \ emission is detected where the \tCO \ emission is strong including main ridge of the Orion A and NGC2023, NGC2024, NGC2068, and NGC2071.
The fact that most of the \CeO($J$ = 2--1) emissions have higher intensity than \CeO($J$ = 1--0) indicates that the region traced by \CeO \ has a temperature and density high enough to excite the molecule to the $J$ = 2 state, and also that the lines are optically thin.
The distribution of \CeO($J$ = 2--1) emission is similar to the distribution of CS \citep{lad91, tat93} emission, which implies \CeO($J$ = 2--1) traces a high density region with $n(\rm H_2)\sim 10^4$ in the cloud.



\subsection{Velocity structure}
\label{sec_velocity}

\begin{figure}
\plotone{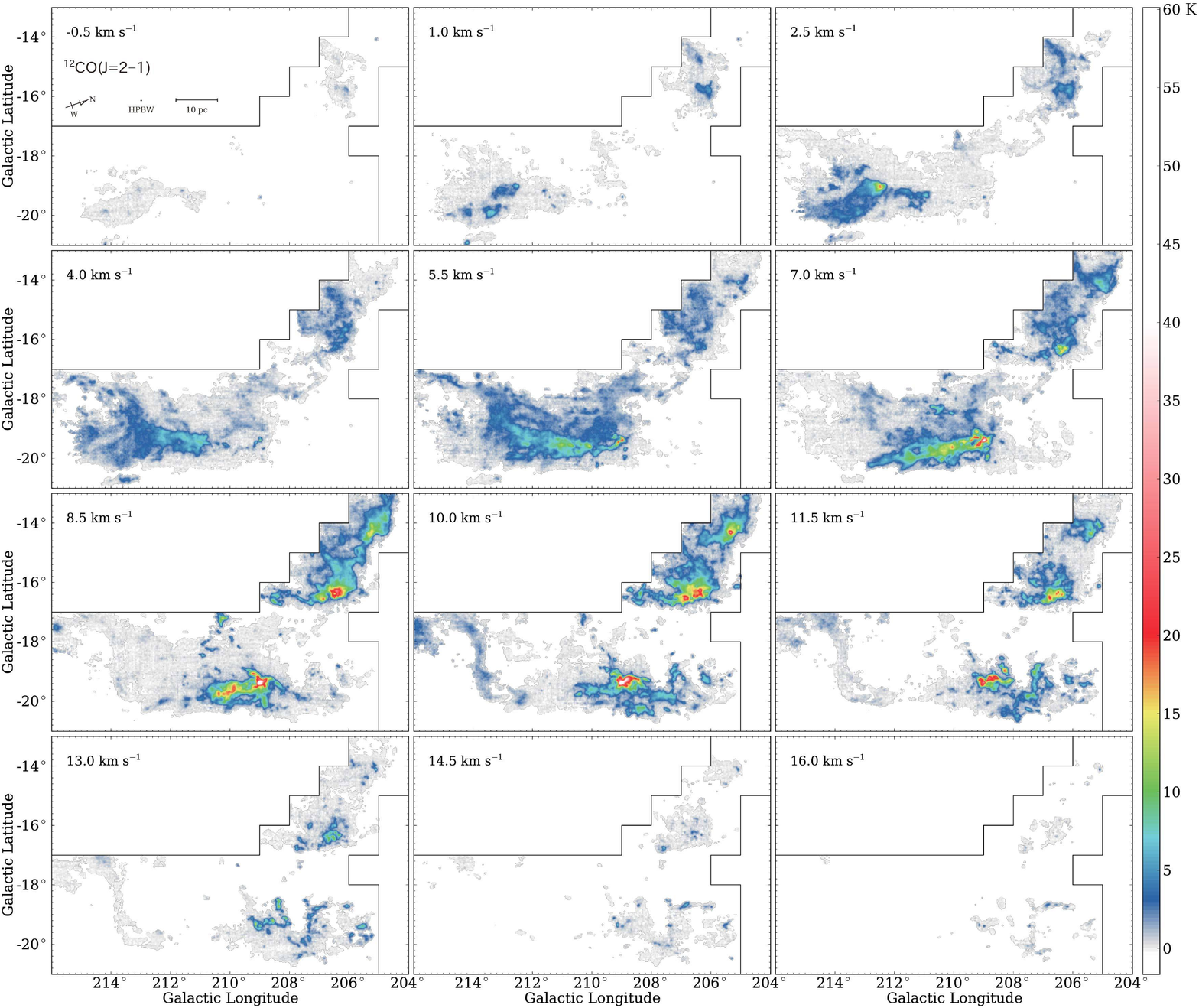}
\caption{
\CO ($J$ = 2--1) velocity channel maps for the velocity range $-0.5 < V_{\rm LSR} < 17.5$ km s$^{-1}$ made at every 1.5 km s$^{-1}$.
The start velocity for the integration is indicated in the top-left corner of each panel.
The moment masked cube was used (see, \S \ref{sec_observation}).
(A color version of this figure is available in the online journal.)
\label{fig_chmap_12CO}}
\end{figure}

\begin{figure}
\plotone{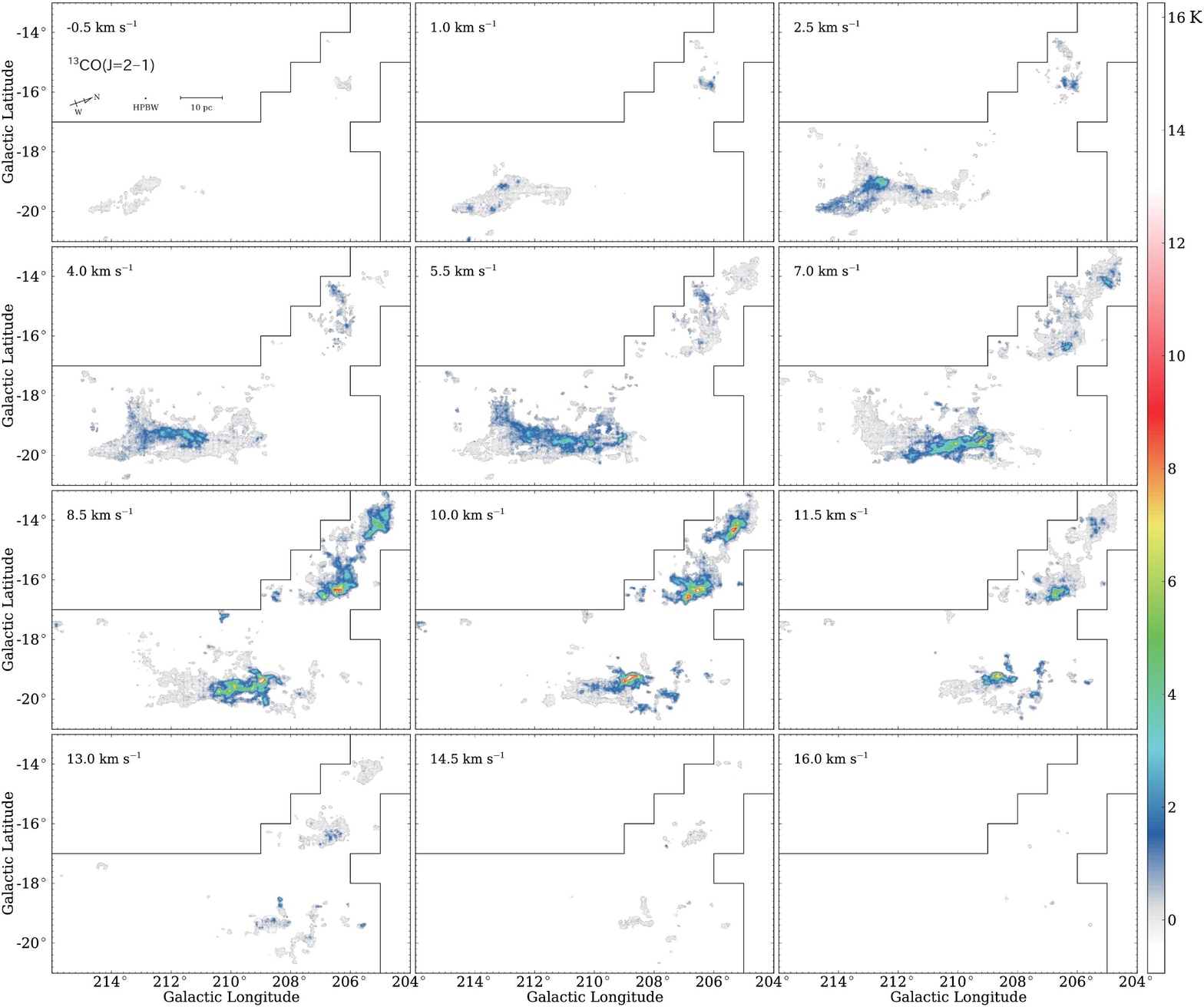}
\caption{
Same as Figure \ref{fig_chmap_12CO}, but for \tCO ($J$ = 2--1).
(A color version of this figure is available in the online journal.)
\label{fig_chmap_13CO}}
\end{figure}

Velocity structures are very complicated in the Orion molecular cloud complex as seen in the velocity channel maps shown in Figure \ref{fig_chmap_12CO} and \ref{fig_chmap_13CO}.
In the Orion A, the main ridge seems to consist of two giant filaments: one is located at $l = 214 \arcdeg$--$211 \arcdeg$ in the velocity range $v_{\rm LSR} = 1.0$--$7.0$ km s$^{-1}$, and the other is located at $l = 212 \arcdeg$--$208 \arcdeg$ in the velocity range $v_{\rm LSR} = 7.0$--$13.0$ km s$^{-1}$.
Both of the filaments consist of many smaller filaments, clumps, and shell-like structures.
The helical filament observed in the velocity range $v_{\rm LSR} = 10.0$--$11.5$ km s$^{-1}$ is the Orion east filament \citep{wil05} which has no physical relation to the Orion main cloud.
The EC is detected at the velocity $v_{\rm LSR} = 5.5$--$8.5$ km s$^{-1}$.
One of the striking features is that the EC consists of many small scale structures (e.g., filaments and clumps) with weak intensities typically $<$ 5 K in the \CO($J$ = 2--1) emission.
The EC clumps have clearly different velocity from the EC which are $v_{\rm LSR} > 8.5$ km s$^{-1}$ with relatively higher intensities typically $> 5$ K in the \CO ($J$ = 2--1) and well-defined boundary.
The Northern clumps are detected in the velocity $v_{\rm LSR} = 10.0$--$16.0$ km s$^{-1}$.
The Northern clumps consist of many small clumps.
There are mainly two velocity components in the Orion B cloud: the lower velocity component corresponding to the 2nd component is found in the velocity range $v_{\rm LSR} = 1.0$--$7.0$ km s$^{-1}$, and the higher velocity component corresponding to the Northern cloud and the Southern cloud is found in the velocity range $v_{\rm LSR} = 7.0$--$16.0$ km s$^{-1}$.
The lower velocity component seems to consist of shells, filaments, and clumps as found for the EC in the Orion A.
On the other hand, the higher velocity component is not very filamentary structure compared with the main ridge in the Orion A.
At the velocity $v_{\rm LSR} > 13.0$ km s$^{-1}$, both of the Orion A and B clouds consist of many small clumps.

\begin{figure}
\plotone{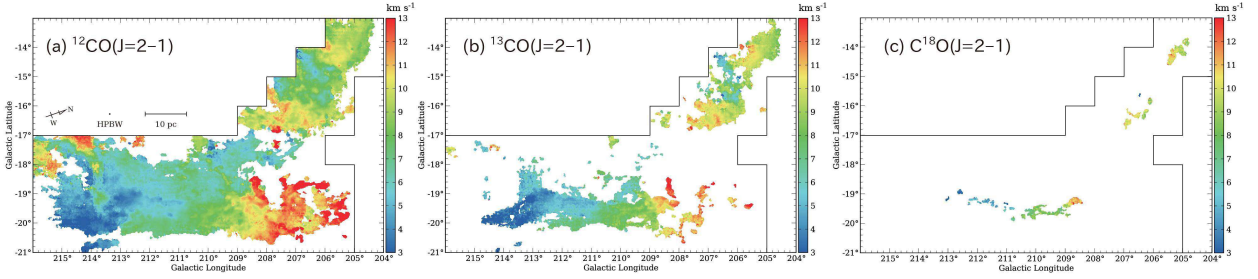}
\caption{
Intensity-weighted mean velocity map in a velocity range from 0 to 20 km s$^{-1}$ for (a)\CO($J$ = 2--1), (b)\tCO($J$ = 2--1), and (c)\CeO($J$ = 2--1).
(A color version of this figure is available in the online journal.)
\label{fig_mom_1}}
\end{figure}

\begin{figure}
\plotone{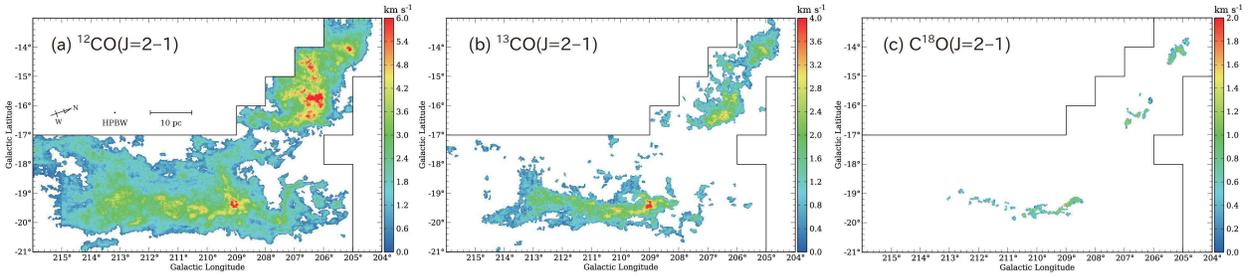}
\caption{
Line width map of (a)\CO($J$ = 2--1), (b)\tCO($J$ = 2--1), and (c)\CeO($J$ = 2--1).
The line widths are obtained by dividing integrated intensity by peak temperature.
(A color version of this figure is available in the online journal.)
\label{fig_mom_2}}
\end{figure}

Figure \ref{fig_mom_1} shows the intensity-weighted mean velocity maps.
The \CO \ and \tCO \ maps exhibit quite similar velocity distribution.
In the Orion A, the main ridge has a velocity gradient while the EC has no velocity change.
The Orion east filament is seen as the high velocity components in the north-east side around $l = 212 \arcdeg$--$216 \arcdeg$ of the Orion A.
In the Orion B, it is clear that the cloud consists of mainly two different velocity components also as seen in Figure \ref{fig_chmap_12CO}.

The line width maps obtained by dividing the integrated intensity by the peak temperature are shown in Figure \ref{fig_mom_2}.
In the Orion A, the line width increases as approaching to the center of the main ridge and as approaching to the Orion KL.
This tendency is more clearly seen in \tCO.
The EC has small line width typically of $<$2 km s$^{-1}$.
In the case of the Orion B cloud, \CO \ emission lines with a very large line width are widely seen, which is explained mainly due to a mixture of some distinct velocity components.
The \tCO \ map seems to trace the velocity structure of the main component of the Orion B as the emission line is optically thinner.
In \tCO, the largest velocity width in the Southern cloud and the Northern cloud are seen toward the NGC2024 and the NGC2071, respectively.

\begin{figure}
\plotone{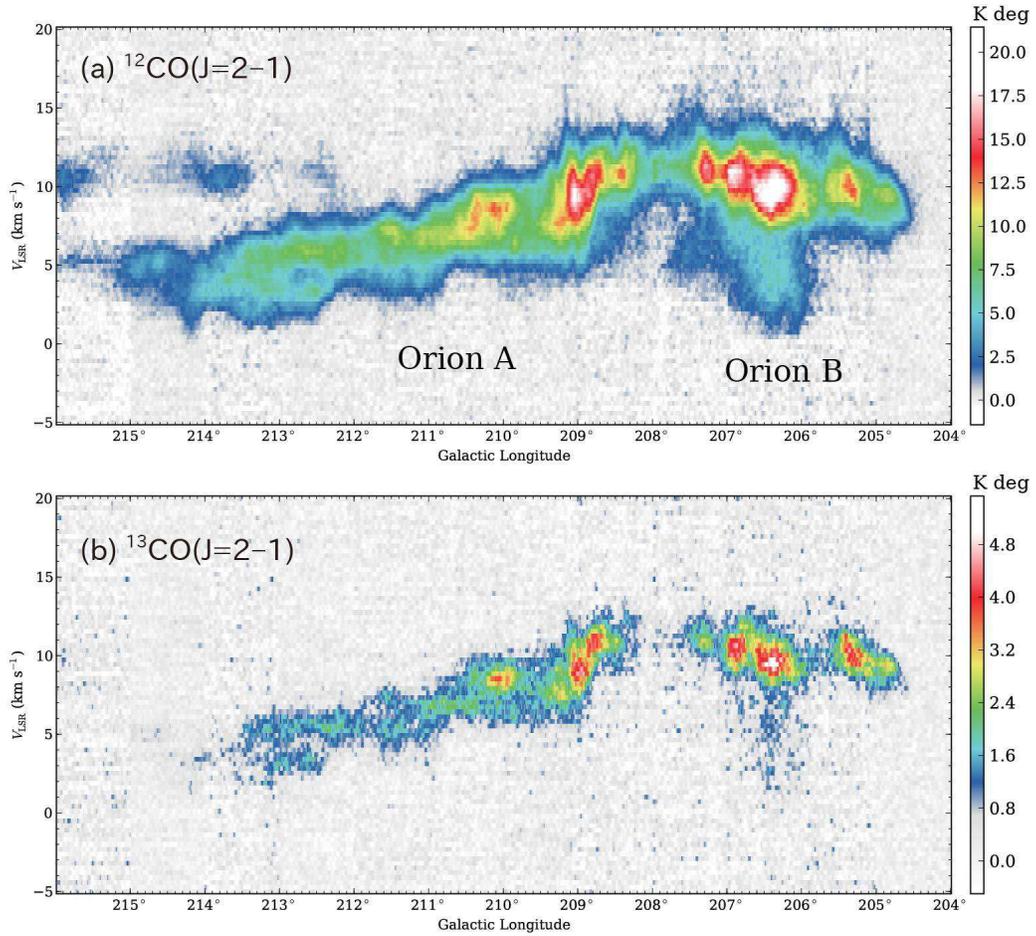}
\caption{
Longitude-velocity (l-v) diagram of the Orion A and B molecular clouds for the emission of (a)\CO ($J$ = 2--1), and (b)\tCO ($J$ = 2--1).
We used spectra in the latitude range between $b = -21\arcdeg$ and $-13\arcdeg$ to produce the diagrams.
(A color version of this figure is available in the online journal.)
\label{fig_pv_lv}}
\end{figure}

\begin{figure}
\plotone{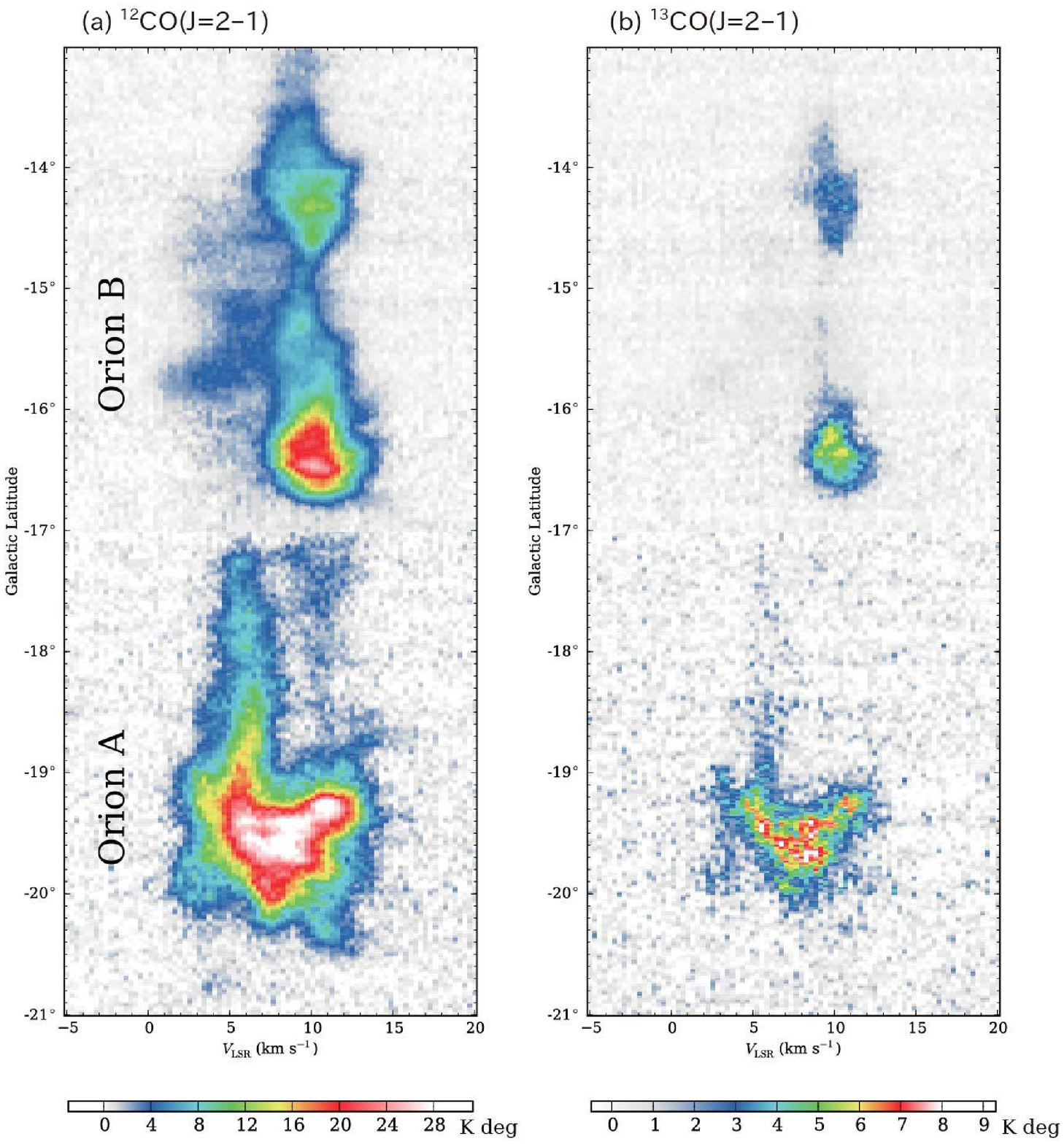}
\caption{
Velocity-latitude (v-b) diagram of the Orion A and B molecular clouds for the emission of (a)\CO ($J$ = 2--1), and (b)\tCO ($J$ = 2--1).
We used spectra in the longitude range between $l = 204\arcdeg$ and $216\arcdeg$ to produce the diagrams.
(A color version of this figure is available in the online journal.)
\label{fig_pv_vb}}
\end{figure}

Figure \ref{fig_pv_lv} shows the longitude-velocity diagrams.
The velocity gradient of the Orion A is also clearly seen in this figure.
It seems to have velocity gradient along longitude also in the Orion B.
The velocity gradients are calculated as $0.15$ and $-0.08$ km s$^{-1}$ pc$^{-1}$, for the Orion A and the Orion B, respectively.
The 2nd component of the Orion B is clearly seen around at $v_{\rm LSR} = 5$ km s$^{-1}$.
The EC of Orion A cannot be identified clearly in the longitude-velocity diagram, because it has the same velocity as the main ridge.

Figure \ref{fig_pv_vb} shows velocity-latitude diagrams.
In the figure, both of the Orion A and B clouds have no velocity gradient.
There is clear boundary between the Orion A and B molecular cloud around $b = -17 \arcdeg$.
In the Orion A, the EC is clearly seen in the velocity around 5 km s$^{-1}$ and the EC clumps are seen with a velocity around 10 km s$^{-1}$.
The extended component of the Orion B is also clearly seen in the velocity-latitude diagram around at $v_{\rm LSR} = 5$ km s$^{-1}$.



\subsection{Line ratios}
\label{sec_ratio}

In this subsection, we derive the intensity ratios of the observed molecular lines to investigate the physical properties of the molecular gas.
We should note that we calculate the ratios without matching the angular resolutions.
Both of the $J$ = 2--1 and $J$ = 1--0 data were originally observed at an angular resolution of $\sim 2\farcm 7$.
However, because the observations by the 1.85-m telescope were carried out in the OTF mode, the resultant angular resolution for the $J$ = 2--1 lines is lowered to $\sim3 \farcm 4$ as stated in Section 2.  
On the other hand, the $J$ = 1--0 observations by the 4-m telescopes were made with an undersampling way, which makes it very difficult to smooth the data exactly to the same angular resolution as that of the $J$ = 2--1 data.
We, therefor, decided not to attempt to standardize the angular resolutions but to use all of the data as that are.
If we observe a point source, the observed intensity would differ by a factor of $\sim 1.5$ due to the difference of the angular resolutions (3\farcm4 or 2\farcm7).
This is the maximum estimate for the possible error in the following analyses arising from the difference of the angular resolutions.
The actual errors should be much smaller, because the CO emission lines are spatially extended as shown in the previous subsections.
In this paper, we neglected all the pixels where the intensities of each line are lower than 3 $\sigma$ noise level when deriving the line ratios.


\subsubsection{Intensity ratio of \CO (J = 2--1)/\CO(J = 1--0)}
\label{Section:Results:Ratio:R12}

\begin{figure}
\plotone{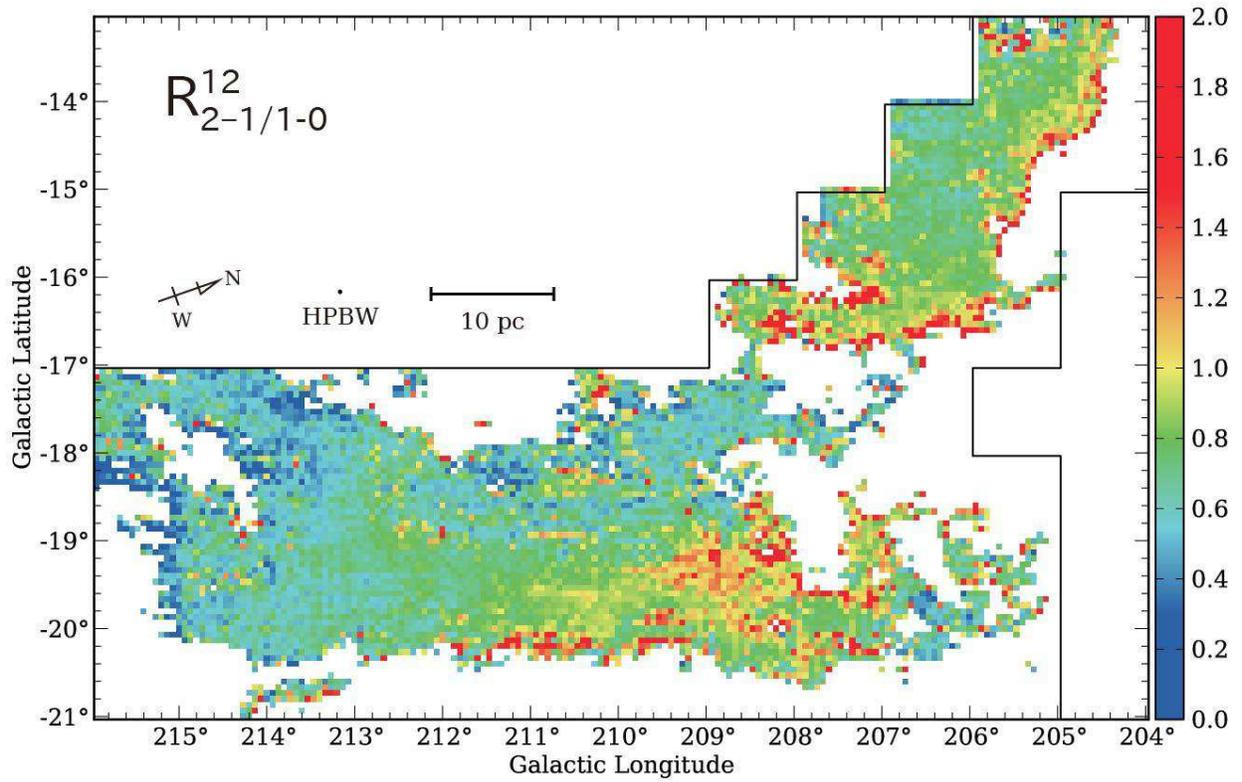}
\caption{
Distribution of the \CO($J$ = 2--1)/\CO($J$ = 1--0) intensity ratio.
The area indicated by the solid line denotes the field observed with the 1.85-m telescope.
(A color version of this figure is available in the online journal.)
\label{fig_ratio_12}}
\end{figure}

Figure \ref{fig_ratio_12} shows the distribution of the \CO ($J$ = 2--1)/\CO ($J$ = 1--0) intensity ratio (hereafter, $R^{12}_{2-1/1-0}$).
In general, the ratio gets close to unity if the emission is quite optically thick, and it reflects the excitation temperature of the region if they are optically thin.
The overall tendency in the figure is similar to that of \citet{sak94}; the ratio is approximately unity along the main ridge of the clouds and decreases down to 0.5 in the peripheral regions. 
The present data reveal the ratio even in much lower intensity regions compared with \citet{sak94} mainly because the present $J$ = 2--1 observations are more sensitive.  

The maximum ratios of $R^{12}_{2-1/1-0}$ are observed toward the cloud boundary near NGC1977 and the lower part of the main ridge around $l = 209 \arcdeg$--$211\arcdeg$ for the Orion A.
For the Orion B, the maximum ratios are observed toward the western side of the main cloud.
In these regions, the ratio becomes higher than 1.5.
The high ratio indicates that the \CO \ lines are optically thin, and that the gas is dense and warm enough to excite to the $J$ = 2 level.
This suggests the interaction of the molecular clouds with the stellar winds and the radiation from the surrounding massive stars.
Relatively high ratio ($\sim 1.3$) is observed near the Orion KL, indicating that the region is also affected by the star clusters in M42 including the Trapezium.

In the Orion A main ridge, a gradient of the ratio is observed, which is previously discovered with \citet{sak94}.
The ratio has the local peaks near L1641N and L1641S which are well-known star forming regions associated with the shell-like structures \citep{hey92}.
The EC is observed as low ratio ($\sim0.5$) while the ratio of the EC clumps are relatively high ($\sim0.8$).
The Northern clumps in the Orion A, the ratio is relatively high especially near the Orion KL.
The ratio of the Orion B is relatively high ($\sim0.8$) except for the 2nd component.


\subsubsection{Intensity ratio of \tCO(J = 2--1)/\tCO(J = 1--0)}

\begin{figure}
\plotone{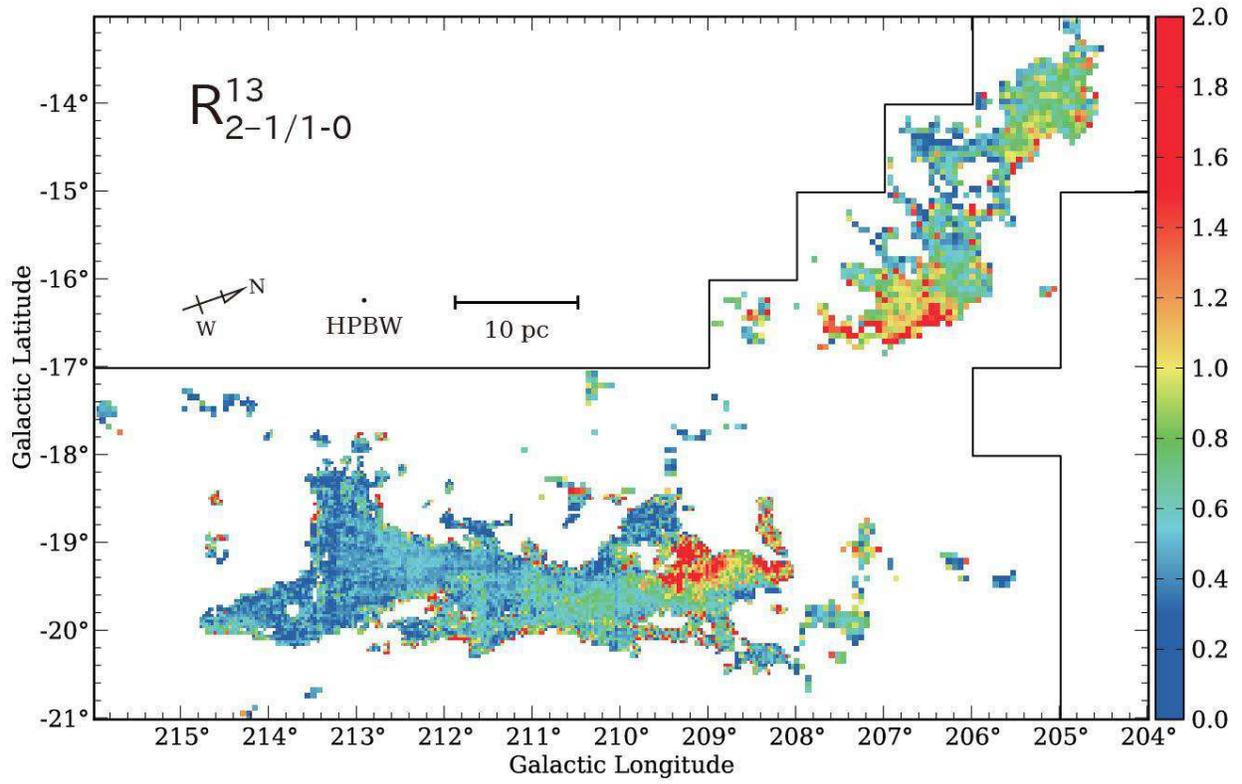}
\caption{
Distribution of the \tCO($J$ = 2--1)/\tCO($J$ = 1--0) intensity ratio.
The area indicated by the solid line denotes the field observed with the 1.85-m telescope.
(A color version of this figure is available in the online journal.)
\label{fig_ratio_13}}
\end{figure}

Figure \ref{fig_ratio_13} shows the distribution of the \tCO($J$ = 2--1)/\tCO($J$ = 1--0) intensity ratio (hereafter, \Rul).
This ratio reflects both the kinematic temperature and density of the gas because of the small optical depth of the \tCO \ emission lines.
The large scale tendency is similar to that of the ratio of \CO \ while the dynamic range is larger.
The maximum ratio in the Orion A is observed toward the cloud boundary near the Orion KL with a ratio of $\sim2$.
The gradient seen in $R^{12}_{2-1/1-0}$ is seen also in \Rul.
We found that the ratio is $\sim0.8$ in the region near L1641N and is $0.3$--$0.5$ in the region at $l > 211.5^\circ$.
Some of the EC clumps and the Northern clumps are detected with the ratio of $\sim0.8$.
In the Orion B clouds, the maximum ratio ($>1.5$) is observed in the western side of NGC2024.
Other clouds in the Orion B observed to be relatively high ratio of $\sim0.9$.


\subsubsection{Intensity ratio of \tCO(J = 2--1)/\CO(J = 2--1)}

\begin{figure}
\plotone{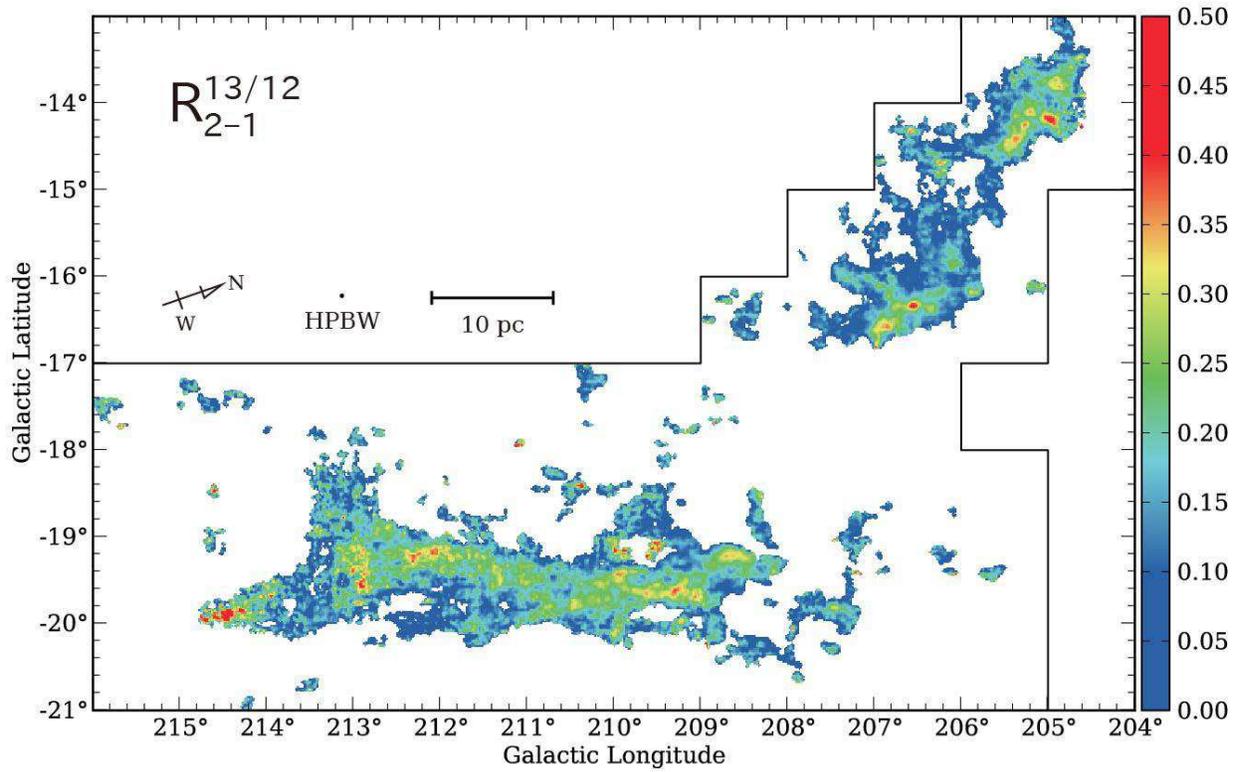}
\caption{
Distribution of the \tCO($J$ = 2--1)/\CO($J$ = 2--1) intensity ratio.
The area indicated by the solid line denotes the field observed with the 1.85-m telescope.
(A color version of this figure is available in the online journal.)
\label{fig_ratio_21}}
\end{figure}

Figure \ref{fig_ratio_21} shows the distribution of the \tCO($J$ = 2--1)/\CO($J$ = 2--1) intensity ratio (hereafter, \Rtt). 
The ratio roughly reflects the column density when the excitation temperatures are the same for both of the lines.
Due to the photon trapping effect, the ratio is also sensitive to local density where the \tCO ($J$ = 2--1) is sub-thermally excited and the \CO($J$ = 2--1) is optically thick.
The ratio is also affected by the abundance variation which mainly reflects the intensity of the interstellar radiation field in the massive star forming region (e.g., \citealt{rip13}).
The distribution of \Rtt \ is somewhat different from the intensity distributions of the \tCO($J$ = 2--1) and also of \CO($J$ = 2--1).  
In Orion A, the ratio is nearly constant from the north to south all along $b \sim -19 \fdg 5$, although the intensity distribution of \tCO($J$ = 2--1) and \CO($J$ = 2--1) are strongest at the northern edge, decreasing to the southern edge.  
In Orion B, the ratio is stronger around the Northern cloud than the Southern cloud, although the tendency is opposite to the intensity distribution of \tCO($J$ = 2--1) and \CO($J$ = 2--1).



\section{Analyses}



\subsection{Comparing column densities and masses derived from the different observed lines}
\label{sec_mass}

\begin{deluxetable}{lrrrrrrrrrrrr}
\tabletypesize{\scriptsize}
\rotate
\tablecaption{
Observed line luminosities and luminosity ratios
\label{table_luminosity}}
\tablehead{
\multicolumn{2}{c}{Source} &
\colhead{$L_{2-1}^{12}$} & 
\colhead{$L_{2-1}^{13}$} & 
\colhead{$L_{2-1}^{18}$} & 
\colhead{$L_{1-0}^{12}$} & 
\colhead{$L_{1-0}^{13}$} & 
\colhead{$L_{1-0}^{18}$} & 
\colhead{$R^{12}_{2-1/1-0}$} & 
\colhead{\Rul} &
\colhead{$R^{18}_{2-1/1-0}$} & 
\colhead{\Rtt} &
\colhead{$R^{13/12}_{1-0}$} \\

\multicolumn{2}{c}{(1)} &
\colhead{(2)} &
\colhead{(3)} &
\colhead{(4)} &
\colhead{(5)} &
\colhead{(6)} &
\colhead{(7)} &
\colhead{(8)} &
\colhead{(9)} &
\colhead{(10)} &
\colhead{(11)} &
\colhead{(12)}
}
\startdata
Orion &  & 23200 & 2930 &  88 & 29500 & 4420 & 158 & 0.79 & 0.66 & 0.56 & 0.13 & 0.15 \\
Orion A &    & 14400 & 1830 &  48 & 19000 & 2890 & 102 & 0.76 & 0.63 & 0.48 & 0.13 & 0.15 \\
\nodata & A1 &  4200 &  413 &   3 &  6950 &  954 &  28 & 0.60 & 0.43 & 0.13 & 0.10 & 0.14 \\
\nodata & A2 & 10200 & 1420 &  45 & 12000 & 1940 &  74 & 0.85 & 0.73 & 0.61 & 0.14 & 0.16 \\
Orion B &    &  8790 & 1100 &  39 & 10500 & 1530 &  56 & 0.84 & 0.72 & 0.70 & 0.12 & 0.15 \\
\nodata & B1 &  5760 &  691 &  18 &  6350 &  830 &  27 & 0.91 & 0.83 & 0.67 & 0.12 & 0.13 \\
\nodata & B2 &  3030 &  408 &  20 &  3730 &  661 &  28 & 0.81 & 0.62 & 0.72 & 0.13 & 0.18 \\
\enddata
\tablecomments{
Col. (1): Source name.
Cols. (2)--(4): Total luminosity of the \CO, \tCO, and \CeO($J$ = 2--1), respectively in K km s$^{-1}$ pc$^2$.
Cols. (5)--(7): Total luminosity of the \CO, \tCO, and \CeO($J$ = 1--0), respectively in K km s$^{-1}$ pc$^2$.
Cols. (8)--(12): Luminosity ratios of the 
$R^{12}_{2-1/1-0} = L_{2-1}^{12} / L_{1-0}^{12}$, 
\Rul$ = L_{2-1}^{13} / L_{1-0}^{13}$, 
$R^{18}_{2-1/1-0} = L_{2-1}^{18} / L_{1-0}^{18}$, 
\Rtt$ = L_{2-1}^{13} / L_{2-1}^{13}$, and 
$R^{13/12}_{1-0} = L_{1-0}^{13} / L_{1-0}^{13}$, respectively.
}
\end{deluxetable}

\begin{deluxetable}{lrrrrrrrrrrr}
\tabletypesize{\scriptsize}
\rotate
\tablecaption{
Averaged column densities and column density ratios
\label{table_column}}
\tablehead{
\multicolumn{2}{c}{Source} &
\colhead{$N_{\rm X}^{1-0}$} & 
\colhead{$N_{\rm LTE}^{13,2-1}$} & 
\colhead{$N_{\rm LTE}^{13,1-0}$} & 
\colhead{$R^{N}_{\rm 13/12}$} &
\colhead{$R^{N}_{\rm LTE,13}$} &
\colhead{$N_{\rm LTE}^{18,2-1}$} & 
\colhead{$N_{\rm LTE}^{18,1-0}$} & 
\colhead{$R^{N}_{\rm 18/12}$} &
\colhead{$R^{N}_{\rm LTE,18}$} \\

\multicolumn{2}{c}{(1)} &
\colhead{(2)} & 
\colhead{(3)} & 
\colhead{(4)} & 
\colhead{(5)} & 
\colhead{(6)} & 
\colhead{(7)} & 
\colhead{(8)} & 
\colhead{(9)} & 
\colhead{(10)} 
}
\startdata
Orion   &    & 18.9 & 20.7 & 22.4 & 1.19 & 0.92 & 62.8 & 44.2 & 2.34 & 1.42 \\
Orion A &    & 15.6 & 19.4 & 20.8 & 1.34 & 0.93 & 61.8 & 44.6 & 2.87 & 1.38 \\
\nodata & A1 & 15.3 & 11.3 & 16.9 & 1.11 & 0.67 & 38.1 & 32.1 & 2.10 & 1.19 \\
\nodata & A2 & 15.8 & 24.1 & 23.2 & 1.47 & 1.04 & 64.9 & 54.4 & 3.45 & 1.19 \\
Orion B &    & 31.4 & 23.3 & 26.4 & 0.84 & 0.88 & 64.2 & 43.4 & 1.38 & 1.48 \\
\nodata & B1 & 37.3 & 25.6 & 28.4 & 0.76 & 0.90 & 74.1 & 47.6 & 1.28 & 1.56 \\
\nodata & B2 & 27.7 & 20.3 & 24.3 & 0.88 & 0.83 & 57.2 & 39.6 & 1.43 & 1.45 \\
\enddata
\tablecomments{
Col. (1): Source name. 
Col. (2): Averaged column density of the H$_2$ derived from \CO($J$ = 1--0) in 10$^{20}$ cm$^{-2}$.
Cols. (3) and (4): Averaged column density of the H$_2$ derived from \tCO($J$ = 2--1) and \tCO($J$ = 1--0), respectively in 10$^{20}$ cm$^{-2}$.
Col. (5): Column density ratio of $R^{N}_{\rm 13/12} = N_{\rm LTE}^{13,1-0} / N_{\rm X}^{1-0}$.
Col. (6): Column density ratio of $R^{N}_{\rm LTE,13} = N_{\rm LTE}^{13,2-1} / N_{\rm LTE}^{13,1-0}$.
Cols. (7) and (8): Averaged column density of the H$_2$ derived from \CeO($J$ = 2--1) and \CeO($J$ = 1--0), respectively in 10$^{20}$ cm$^{-2}$.
Col. (9): Column density ratio of $R^{N}_{\rm 18/12} = N_{\rm LTE}^{18,1-0} / N_{\rm X}^{1-0}$.
Col. (10): Column density ratio of $R^{N}_{\rm LTE,18} = N_{\rm LTE}^{18,2-1} / N_{\rm LTE}^{18,1-0}$.
}
\end{deluxetable}

\begin{deluxetable}{lrrrrrrrrrrr}
\tabletypesize{\scriptsize}
\rotate
\tablecaption{
Total masses and mass ratios
\label{table_mass}}
\tablehead{
\multicolumn{2}{c}{Source} &
\colhead{$M_{\rm X}^{1-0}$} & 
\colhead{$M_{\rm LTE}^{13,2-1}$} & 
\colhead{$M_{\rm LTE}^{13,1-0}$} & 
\colhead{$R^{M}_{\rm 13/12}$} &
\colhead{$R^{M}_{\rm LTE,13}$} &
\colhead{$M_{\rm LTE}^{18,2-1}$} & 
\colhead{$M_{\rm LTE}^{18,1-0}$} & 
\colhead{$R^{M}_{\rm 18/12}$} &
\colhead{$R^{M}_{\rm LTE,18}$} \\

\multicolumn{2}{c}{(1)} &
\colhead{(2)} & 
\colhead{(3)} & 
\colhead{(4)} & 
\colhead{(5)} & 
\colhead{(6)} & 
\colhead{(7)} & 
\colhead{(8)} & 
\colhead{(9)} & 
\colhead{(10)} 
}
\startdata
Orion   &    & 110.4 &  28.2 &  90.5 & 0.82 & 0.31 &   6.9 &  45.3 & 0.41 & 0.15 \\
Orion A &    &  70.6 &  17.5 &  59.8 & 0.85 & 0.29 &   3.9 &  28.1 & 0.40 & 0.14 \\
\nodata & A1 &  25.8 &   3.8 &  18.2 & 0.71 & 0.21 &   0.3 &   8.9 & 0.34 & 0.03 \\
\nodata & A2 &  44.8 &  13.7 &  41.6 & 0.93 & 0.33 &   3.6 &  19.2 & 0.43 & 0.19 \\
Orion B &    & 39.7 & 10.7 &  30.7 & 0.77 & 0.35 &   3.0 &  17.2 & 0.43 & 0.18 \\
\nodata & B1 &  24.0 &   6.7 &  18.2 & 0.76 & 0.37 &   1.4 &   9.0 & 0.37 & 0.16 \\
\nodata & B2 &  14.2 &   4.0 &  12.5 & 0.88 & 0.32 &   1.6 &   8.2 & 0.58 & 0.19 \\
\enddata
\tablecomments{
Col. (1): Source name. 
Col. (2): Total molecular cloud mass derived from \CO($J$ = 1--0) in $10^3 M_{\sun}$.
Cols. (3) and (4): Total molecular cloud mass derived from \tCO($J$ = 2--1) and \tCO($J$ = 1--0), respectively in $10^3 M_{\sun}$.
Col. (5): Mass ratio of $R^{M}_{\rm 13/12} = M_{\rm LTE}^{13,1-0} / M_{\rm X}^{1-0}$.
Col. (6): Mass ratio of $R^{M}_{\rm LTE,13} = M_{\rm LTE}^{13,2-1} / M_{\rm LTE}^{13,1-0}$.
Cols. (7) and (8): Total molecular cloud mass derived from \CeO($J$ = 2--1) and \CeO($J$ = 1--0), respectively in $10^3 M_{\sun}$.
Col. (9): Mass ratio of $R^{M}_{\rm 18/12} = M_{\rm LTE}^{18,1-0} / M_{\rm X}^{1-0}$.
Col. (10): Mass ratio of $R^{M}_{\rm LTE,18} = M_{\rm LTE}^{18,2-1} / M_{\rm LTE}^{18,1-0}$.
}
\end{deluxetable}

Column densities of the molecular gas are often derived by assuming the X-factor, which is a conversion factor from line intensities to column densities for optically thick lines, and by assuming the Local Thermodynamic Equilibrium (LTE) for optically thin lines.  
In this subsection, we derive the column densities and the masses
by using the assumptions in the above, and discuss the cause of their differences.  
In order to investigate the difference depending on the environments in terms of the star formation activity, we divide the observed area into four regions, i.e., Orion A-1, Orion A-2, Orion B-1, and Orion B-2 (see Figure \ref{fig_finding}).
Orion A-1 is a part of Orion A at $l > 211\arcdeg$, and it includes no massive star formation site.  
Orion A-2 is the region at $l < 211\arcdeg$ where the massive star formation is taking place.
Orion B-1 is a part of Orion B at $b < 15\arcdeg$ corresponding to the Southern cloud as introduced in \S \ref{sec_12CO},
and Orion B-2 is the region at $b > 15\arcdeg$ corresponding to the Northern cloud.


\subsubsection{Line luminosities}

We summarize the luminosity of the observed emission lines and their ratios in Table \ref{table_luminosity}.
To derive the intensity, we integrated the observed emission lines over the surface areas of each subregion.
The ratio of $J$ = 2--1/$J$ = 1--0 is different depending on the isotopes.
The $R^{12}_{2-1/1-0}$ is the highest $\sim$0.6--0.9 and the $R^{18}_{2-1/1-0}$ is the lowest $\sim$0.1--0.7.
Especially in the A1 subregion, $R^{18}_{2-1/1-0}$ is very low compared with $R^{12}_{2-1/1-0}$ by a factor of 3.
The ratios of \tCO /\CO \ show similar tendency both in $J$ = 2--1 and $J$ = 1--0.
The A2 subregion is higher than the A1 subregion and the B2 subregion is higher than the B1 subregion.


\subsubsection{Column densities}

The X-factor, which converts from \CO($J$ = 1--0) line intensities to the column densities of molecular hydrogen, has been derived by comparing the intensities with other tracers of mass, such as virial masses (e.g., \citealt{sol87}), proton masses from gamma-ray observations (e.g., \citealt{blo86}), and dust observations(e.g., \citealt{dam01}). 
For the Galactic clouds, the X-factor is derived to be approximately $1.8 \times 10^{20}$ cm$^{-2}$ K$^{-1}$ km$^{-1}$ s \citep{dam01}, and we use this value in this paper.  
The averaged column densities derived with the X-factor is
$N^{1-0}_{\rm X}({\rm H_2}) = 18.9 \times 10^{20}$ cm$^{-2}$ 
.
We also derived the averaged column densities for each subregion and summarized them in Table \ref{table_column}.

$J$ = 1--0 transition of the \tCO \ and \CeO \ have been often used to derive the column density under the assumption of the LTE (e.g., \citealt{dic78,pin10}), because the Einstein's A coefficient is small, and thus the critical density for the excitation is low.  
The $J$ = 2--1 transitions have higher critical densities for the excitation, and they can be sub-thermally excited in lower-density regions.
In the analyses, we apply the LTE assumption for all of the transition lines, and discuss the cause of the differences of the derived properties.
Furthermore, we use the peak brightness temperature of each \CO \ transition line for the estimation of the excitation temperature of the \tCO \ and \CeO \ transitions.
Assuming the LTE, the excitation temperature $T_{\rm ex}$ is derived from the peak brightness temperature of \CO \ line, $T_{\rm peak}$, as
\begin{equation}
T_{\rm ex}^{1-0} = 5.53 \left\{ \ln \left[ 1+ \frac{5.53}{T_{\rm peak}^{12,1-0} + 0.84} \right] \right\}^{-1}
\end{equation}
\begin{equation}
T_{\rm ex}^{2-1} = 11.06 \left\{ \ln \left[ 1+ \frac{11.06}{T_{\rm peak}^{12,2-1} + 0.19} \right] \right\}^{-1}.
\end{equation}
Using the excitation temperature, the optical depths of the \tCO \ and \CeO \ emissions lines are derived from the brightness temperature, $T_{\rm mb}(v)$, 
\begin{equation}
\tau_{J=1}^{13}(v) = - \ln \left\{ 1 - \frac{T_{\rm mb}^{13,1-0}(v)}{5.29} \left[ \frac{1}{\exp(5.29/T_{\rm ex}) - 1} - 0.17 \right]^{-1} \right\}
\end{equation}
\begin{equation}
\tau_{J=1}^{18}(v) = - \ln \left\{ 1 - \frac{T_{\rm mb}^{18,1-0}(v)}{5.27} \left[ \frac{1}{\exp(5.27/T_{\rm ex}) - 1} - 0.17 \right]^{-1} \right\}
\end{equation}
\begin{equation}
\tau_{J=2}^{13}(v) = - \ln \left\{ 1 - \frac{T_{\rm mb}^{13,2-1}(v)}{10.58} \left[ \frac{1}{\exp(10.58/T_{\rm ex}) - 1} - 0.02 \right]^{-1} \right\}
\end{equation}
\begin{equation}
\tau_{J=2}^{18}(v) = - \ln \left\{ 1 - \frac{T_{\rm mb}^{18,2-1}(v)}{10.54} \left[ \frac{1}{\exp(10.54/T_{\rm ex}) - 1} - 0.02 \right]^{-1} \right\}.
\end{equation}
The column densities of \tCO \ and \CeO \ in the upper state, $N_u$, are derived by the following equations,
\begin{equation}
N_{J=1}^{13} = 1.98 \times 10^{16} \left[ \exp \left( \frac{5.29}{T_{\rm ex}} \right) - 1 \right]^{-1} \int \tau_{J=1}^{13}(v) dv
\end{equation}
\begin{equation}
N_{J=1}^{18} = 1.97 \times 10^{16} \left[ \exp \left( \frac{5.27}{T_{\rm ex}} \right) - 1 \right]^{-1} \int \tau_{J=1}^{18}(v) dv
\end{equation}
\begin{equation}
N_{J=2}^{13} = 1.65 \times 10^{16} \left[ \exp \left( \frac{10.58}{T_{\rm ex}} \right) - 1 \right]^{-1} \int \tau_{J=2}^{13}(v) dv
\end{equation}
\begin{equation}
N_{J=2}^{18} = 1.64 \times 10^{16} \left[ \exp \left( \frac{10.54}{T_{\rm ex}} \right) - 1 \right]^{-1} \int \tau_{J=2}^{18}(v) dv .
\end{equation}
Assuming the LTE, the column density of the rotational state of $J$ is related to the total CO column density as
\begin{equation}
N_{\rm total}({\rm CO}) = N_{J} \frac{Z}{2J+1} \exp \left[ \frac{h B_0 J (J+1)}{k T_{\rm ex}} \right]
\end{equation}
where $B_0$ is the rotational constant of the CO isotopologues, $B_0 = 5.51 \times 10^{10}$ s$^{-1}$ and $5.49 \times 10^{10}$ s$^{-1}$ for \tCO \ and \CeO, respectively.
$Z$ is the partition function which is given by
\begin{equation}
Z = \sum_{J=0}^{\infty} (2J + 1) \exp \left[ - \frac{h B_0 J (J+1)}{k T_{\rm ex}} \right] .
\end{equation}
The column density of the molecular gas, $N(\rm H_2)$, is derived by
\begin{equation}
N({\rm H_2}) = X N_{\rm total}({\rm CO})
\end{equation}
where $X$ is the isotopic abundance ratio of the CO isotopologues relative to H$_2$.
We adopt $X[{\rm ^{13}CO}] = 7.1 \times 10^5$ and $X[{\rm C^{18}O}] = 5.9 \times 10^6$ \citep{fre82}.

The derived averaged column densities over the whole observed area from \tCO \ and \CeO \ of $J$ = 2--1 and $J$ = 1--0 are 
$N^{13,2-1}_{\rm LTE} = 20.7 \times 10^{20}$ cm$^{-2}$,
$N^{13,1-0}_{\rm LTE} = 22.4 \times 10^{20}$ cm$^{-2}$,
$N^{18,2-1}_{\rm LTE} = 62.8 \times 10^{20}$ cm$^{-2}$, and
$N^{18,1-0}_{\rm LTE} = 44.2 \times 10^{20}$ cm$^{-2}$, respectively.
The derived column densities are summarized in Table \ref{table_column}.

We note here that the \tCO \ and \CeO \ abundances can change depending on the surrounding environments although we assume the uniform distribution of the abundances throughout the clouds.
The abundances seem to depend on the self-shielding and the star formation activities, and the values for the \tCO \ ranges mostly within [\tCO]/[H$_2$] = $1$--$3.5 \times 10^{-6}$ (e.g., \citealt{dic78, fre82, lad94, har04, pin08, pin10, sim11, rip13}), which affect the estimations of the mass and the column densities.

The column densities derived from \CO \ are similar to that of \tCO \ while the \CeO \ show significant higher averaged column densities.
This indicates the \CeO \ emission traces higher column density region than \CO \ and \tCO, probably due to the photodissociation and chemical fractionation of the species (e.g., \citealt{war96}).
Another possibility is that the abundance ratio of \CeO \ in the Orion region is different from those in the other regions measured by Frerking et al. (1982).


\subsubsection{Masses}

The gas mass is calculated from the molecular gas column densities by
\begin{equation}
\left( \frac{M_{\rm gas}}{M_{\sun}} \right) = 
4.05 \times 10^{-1} \mu_{\rm H_2}
\left( \frac{m_{\rm H}}{\rm kg} \right)
\left( \frac{d}{\rm pc} \right)^2
\left( \frac{\Delta l}{\rm arcmin} \right)
\left( \frac{\Delta b}{\rm arcmin} \right)
\left( \frac{N({\rm H_2})}{\rm cm^{-2}} \right)
\end{equation}
where $\mu_{\rm H_2} \sim 2.7$ is the mean molecular weight per H$_2$ molecule, $m_{\rm H}$ is the atomic hydrogen mass, $d$ is the distance, and $\Delta l$ and $\Delta b$ are the pixel size along the galactic coordinates.

The derived gas masses are summarized in Table \ref{table_mass}.
The masses derived from the $J$ = 1--0 are larger than those derived from the $J$ = 2--1 in all the CO isotopes.
The total gas masses derived from \tCO($J$ = 1--0) for four regions are about 70--90\% of those derived from \CO($J$ = 1--0) luminosities.
The ratio of the total masses derived from the two molecular lines are almost uniform not depending on the regions.  
This implies that the optically thick \CO($J$ = 1--0) line is well proportional to the total mass, and if we assume that the mass derived from \tCO($J$ = 1--0) traces the true total mass more reliably, the X factor for \CO($J$ = 1--0) intensity is estimated to be $1.5 \times 10^{20}$ cm$^{-2}$ K$^{-1}$ km$^{-1}$ s.  
The mass derived from \tCO ($J$ = 2--1) is lower than that from \tCO($J$ = 1--0) by a factor of about 3, indicating that the $J$ = 2--1 line is sub-thermally excited.  
Especially toward the A1 subregion, the ratio \tCO ($J$ = 2--1)/\tCO ($J$ = 1--0) is lower than the other regions by a factor of 1.4.
This indicates that the density of the Orion A1 region is lower than other two regions, which is also discussed in the previous sub-subsection.



\subsection{Large velocity gradient analyses}

Molecular lines with different critical densities for the excitation can be used to estimate the density and the temperature of the emitting region.  
For the optically thick molecular lines, we need to include an effect of photon trapping.  
The photon trapping varies the excitation state and depends on the morphology of the cloud. 
For simplicity, we used the large velocity gradient (LVG) approximation method (e.g., \citealt{gol74, sco74}), which assumes a spherically symmetric cloud of uniform density and temperature with a spherically symmetric velocity distribution proportional to the radius and uses a Castor escape probability formalism \citep{cas70}. 
It solves the equations of statistical equilibrium for the fractional population of CO rotational levels at each density and temperature by incorporating the photon escape probability that is effective in the optically thick case.  
The widely used radiative transfer code RADEX also uses the same technique with an ability to choose three different formulations of the escape probability \citep{van07}.
In the non-LTE analyses, intensities of a few lines are compared and used to determine the physical properties of the gas where lines are emitted.
Therefore, the analyses are sensitive to the density similar to the critical densities of the used lines (e.g., \citealt{cas90, beu00, zhu03}).
The selection of the lines for the non-LTE analyses are important.
Recently, the combination of optically thin and thick lines with different transitions are found to be good tracers of the physical properties of the gas and the derived physical properties well reflect the star formation activities and the surrounding environments (e.g., \citealt{mar04, nag07, miz10, min11, tor11, naz12, pen12, fuk14}).
In this paper, we use the \CO($J$=2--1), \tCO($J$=2--1), and \tCO($J$=1--0) lines with the single component LVG analyses.

\begin{figure}
\plotone{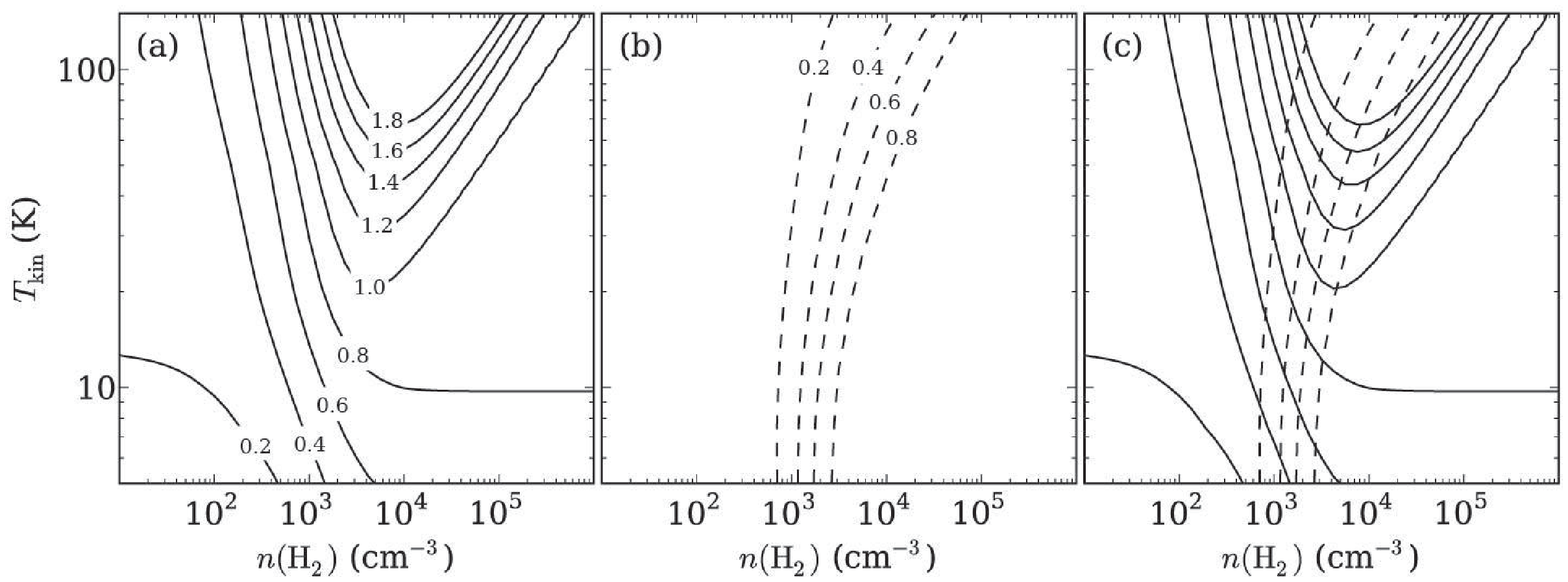} 
\caption{
Contour plots of the calculated line intensity ratio using the LVG analyses.
Contours are the values of (a)$R^{13}_{2-1/1-0}$, (b)$R^{13/12}_{2-1}$, and (c) $R^{13}_{2-1/1-0}$ and $R^{13/12}_{2-1}$.
We assumed, X(\CO) = $1 \times 10^{-4}$ , $dv/dr = 1.0$ km s$^{-1}$ pc$^{-1}$, and the abundance ratio of \CO/\tCO \ = 71.
\label{fig_lvg_contour}}
\end{figure}

We use the intensity ratios, not the absolute intensities, for the analyses to minimize the effect of the beam filling factor when deriving the density and temperature.
This is because the high spatial resolution observations revealed that molecular clouds contain many small-scale structures (see \citealt{rip13, nak12, shi11} for the Orion case), and the beam filling factors are not unity.
\citet{sak94} discussed the effect of the unresolved clumps for their analyses based on the radiative transfer computations of clumps by \citet{gie92}, and suggested that the ratios reflect the physical conditions in constituent clumps and that the beam filling factors may change depending on the regions.

Our analyses includes the lowest 40 rotational levels of the ground vibrational level and uses Einstein A coefficient and ortho/para H$_{2}$ impact rate coefficients of \citet{sch05}. 
The ratio of ortho- to para-H$_{2}$ molecules is calculated by assuming the thermal equilibrium state. 
We performed the calculations for a $^{12}$CO fractional abundance of X($^{12}$CO) = [$^{12}$CO]/[H$_{2}$] = $1\times10^{-4}$ and a $^{12}$CO/$^{13}$CO abundance ratio of 71 \citep{fre82}.  
Another parameter we need for the calculation is the velocity gradient often described as $dv/dr$.  
Figure \ref{fig_lvg_contour} shows contour plots of the LVG analyses by assuming $dv/dr$ to be 1 km s$^{-1}$ pc$^{-1}$ which are derived from the typical line width and the size of the cloud.
Solid and dashed lines show contours of \Rul \ and \Rtt \ ratios, respectively.
The figure indicates that the \Rtt \ ratio basically depends on the density, and the \Rul \ ratio depends on both of the density and temperature.  
The  \Rtt \ dependency comes from the facts that it reflects the optical depth of \tCO($J$ = 2--1) when \CO($J$ = 2--1) is optically thick, and also that less H$_{2}$ density is needed for the collisional excitation of \CO($J$ = 2--1) than the optically thin \tCO($J$ = 2--1) line due to the photon trapping effect of the \CO($J$ = 2--1) line.  
When \tCO \ lines are optically thin, \Rul \ does not depend on the column density or the change of the abundances, reflecting only the local physical properties of the density and temperature because of the absence of the photon trapping effect.
This makes the ratio a good tracer of the physical properties.
\Rul \ is dependent only on the temperature if both of the lines are optically thin and fully thermalized.
This ratio also depends on the density in terms of the different critical density for the excitation.
Roughly speaking, \Rtt \ traces the density, and \Rul \ is larger for higher temperature and density.  
Because these two ratios have different dependence on the density and temperature, we are able to estimate the density and temperature from the intersection in the figure.  

\begin{figure}
\plotone{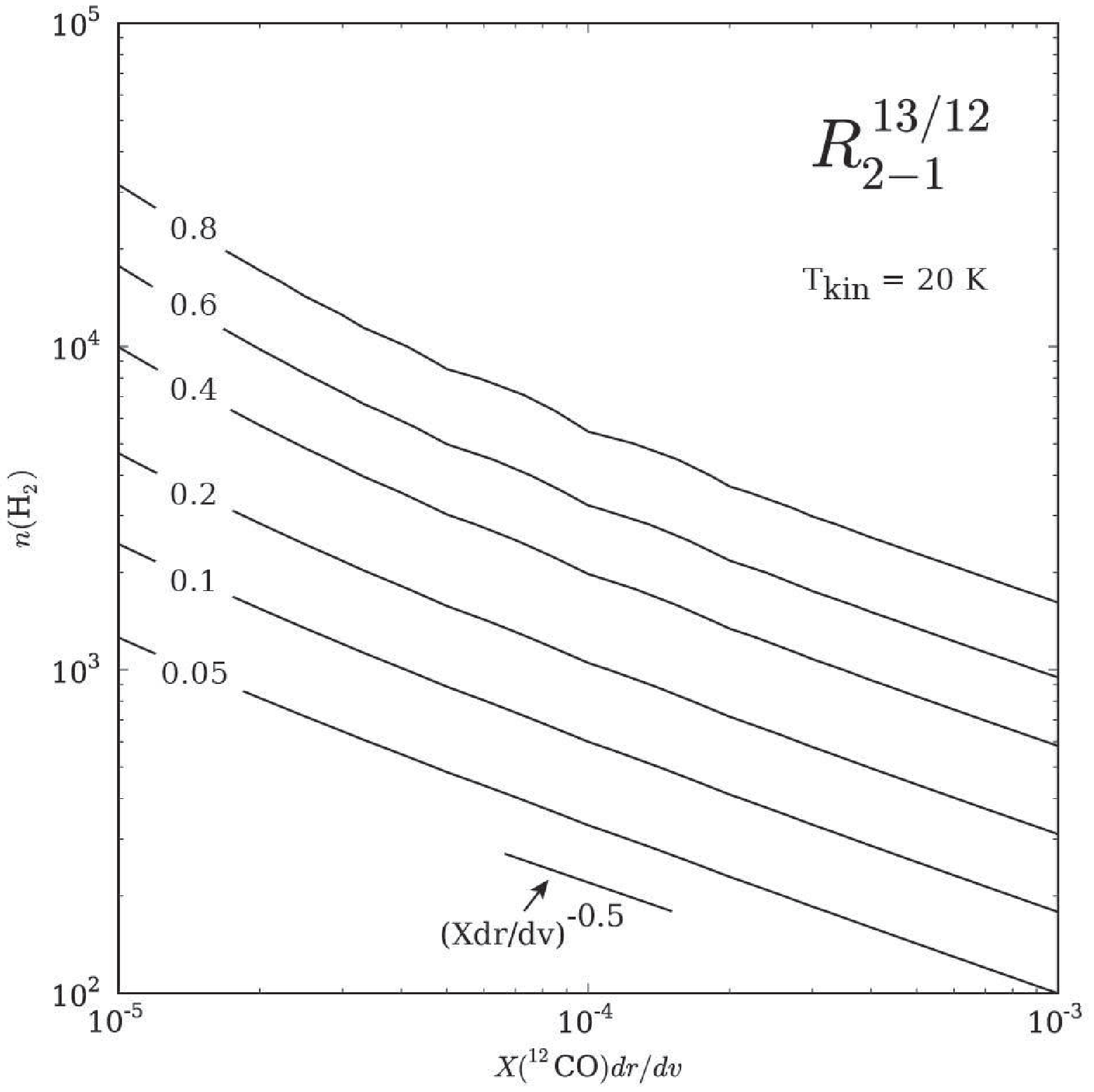} 
\caption{
Contour plots of the calculated line intensity ratio \Rtt \ as a function of hydrogen density and CO fractional abundance per unit velocity gradient for the kinematic temperature of 20 K.
\label{fig_lvg_contour_x}}
\end{figure}

In the present paper, we assumed uniform fractional abundance of CO and $dv/dr$ for the analyses.
The change of the parameters results in the change of the derived $n$(H$_2$), and we estimate the effect here.
The abundance fluctuations of [\tCO]/[H$_2$] range mostly within a factor of 2 as indicated in sub-section 4.1.2.
Especially, [\tCO]/[H$_2$] is observed to be fairly constant in self-shielded clouds where \tCO($J$=1--0) intensity is relatively high \citep{rip13}.
The observed line width should reflect the ratio $dv/dr$.
Figure \ref{fig_mom_2}b indicates that the line width ranges mainly between 1.5 to 3 km s$^{-1}$.
We also calculated the change of the derived $n$(H$_2$) with the different assumptions of $X dr/dv$.  
For a fixed \tCO($J$=2--1)/\CO($J$=2--1) ratio, the derived density is inversely proportional to square root of the assumed $X dr/dv$ (Figure \ref{fig_lvg_contour_x}).
This means that even in a factor of 10 variation in $X dr/dv$ only amounts to a factor of $\sim3$ in density.

Finally, we discuss briefly the effect of the density inhomogeneities on the above analyses.
We assumed here the uniform physical properties in a beam.
However, the assumption may not be realistic because the density can vary along the line of sight.
Because we used the ratios of the line intensities, the derived densities can vary depending on the density distributions.
For example, in the peripheral area, where the lower density gas is extended, the density contrast may not be so high.
In this case, the derived density roughly represents the physical properties of the extended gas.
On the other hand, toward dense core regions, the high density gas is normally surrounded by the lower density gas, and thus the derived density may not reflect the physical properties of the high density gas.
In this case, the derived density can thus vary depending on the density and optical depth of the lower density gas.
Therefore, we need to estimate how large the lower density gas can affect the line intensities although it is not easy because it depends much on the morphology of the clouds.
In the present analyses, we are making use of mainly the sub-thermality of the optically thin \tCO($J$=2--1) line to derive the density.
In a sub-thermal density range, the line intensity depends much on the density.
With the column density fixed, the intensity is almost proportional to the density (e.g., see Figure A2 of \citealt{gin11}), indicating that the higher density gas contributes to the line intensity more than the lower density gas, putting more weight to the higher density gas.
This implies that the derived densities here represent those of the higher density gas rather than those of the surrounding lower density gas.
For example, \citet{sne84} calculated the effect of the lower density gas for CS lines and concluded that the effect is small when the optical depth of foreground density gas is less than 0.5.
\citet{mun86} modeled spherical clouds with $1/R$ and $1/R^{2}$ density dependence and suggested that the calculated line intensities for the optically thin line yield densities that are between the maximum and average densities along the line of sight.

\subsubsection{Deriving physical parameters}

\begin{deluxetable}{lrrrrrrrrrl}
\tabletypesize{\scriptsize}
\tablecaption{
Results of LVG analyses
\label{table_lvg}}
\tablehead{
\colhead{} & 
\colhead{$l$} & 
\colhead{$b$} & 
\colhead{$T^{12}_{2-1}$} & 
\colhead{$T^{13}_{2-1}$} & 
\colhead{$T^{13}_{1-0}$} & 
\colhead{$R^{13/12}_{2-1}$} & 
\colhead{$R^{13}_{2-1/1-0}$} & 
\colhead{$T_{\rm kin}$} & 
\colhead{$n(\rm H_2)$} & 
\colhead{} \\

\colhead{Source} & 
\colhead{(deg)} & 
\colhead{(deg)} & 
\colhead{(K)} & 
\colhead{(K)} & 
\colhead{(K)} & 
\colhead{} & 
\colhead{} & 
\colhead{(K)} & 
\colhead{(cm$^{-3}$)} & 
\colhead{}
}
\startdata
Orion KL & 209.00 & $-19.40$ & 57.8 & 15.6 & 10.1 & 0.27 & 1.54 & 88 & 1800 & \\
OMC-3 & 208.60 & $-19.20$ & 23.8 & 12.2 & 10.9 & 0.51 & 1.12 & 34 & 2200 & \\
L1641-N & 210.07 & $-19.67$ & 19.2 & 5.8 & 7.7 & 0.30 & 0.76 & 21 & 1300 & \\
L1641-S & 212.00 & $-19.33$ & 6.8 & 2.1 & 4.1 & 0.31 & 0.51 & 10 & 1000 & \\
NGC2024 & 206.53 & $-16.33$ & 20.2 & 13.6 & 14.7 & 0.67 & 0.93 & 30 & 2200 & \\
NGC2023 & 206.87 & $-16.53$ & 30.2 & 12.8 & 13.9 & 0.43 & 0.92 & 33 & 2000 & \\
NGC2068 & 205.40 & $-14.33$ & 25.6 & 12.5 & 14.2 & 0.49 & 0.88 & 26 & 1600 & \\
NGC2071 & 205.13 & $-14.13$ & 14.3 & 6.8 & 9.6 & 0.48 & 0.71 & 30 & 1400 & \\
\enddata
\end{deluxetable}

\begin{figure}
\plotone{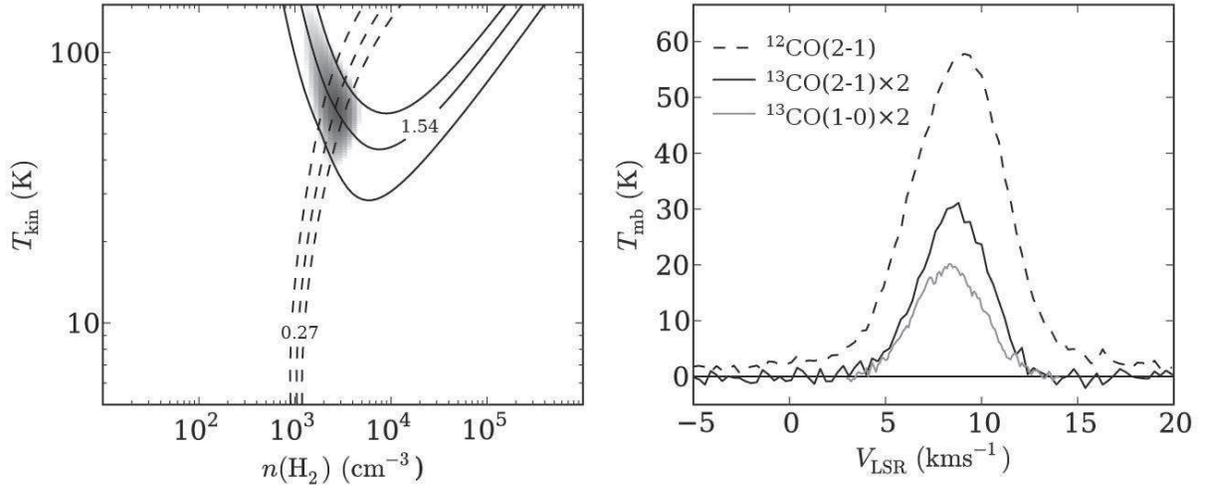}
\caption{
(left)Contour plots of the LVG analyses of the Orion KL region with $Xdr/dv = 1.0 \times 10^{-4} $ pc km$^{-1}$ s.
The vertical axis is kinetic temperature $T_{\rm kin}$, and the horizontal axis is molecular hydrogen density $n(\rm H_2)$.
Solid lines represent \Rul , and dashed lines represent \Rtt with intensity calibration errors of 10\%.
Gray scales show the results of $\chi^2$ test.
(right)Spectra used for the LVG analyses. The dashed line represents \CO($J$ = 2--1), solid black line represents \tCO($J$ = 2--1), and solid gray line represents \tCO($J$ = 1--0).
The \tCO \ are scaled up by a factor of 2.
\label{fig_lvg_orikl}}
\end{figure}

\begin{figure}
\plotone{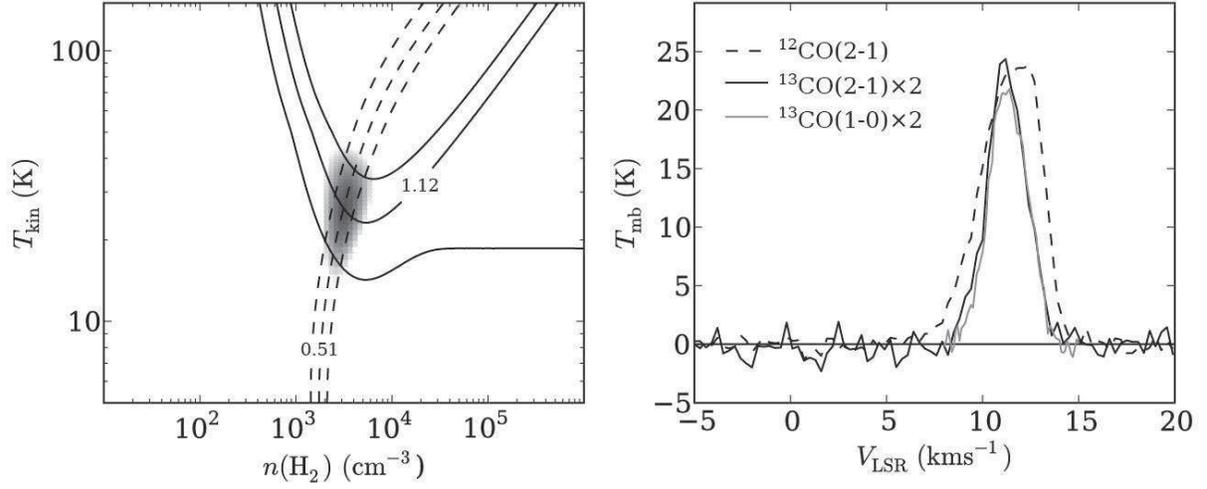}
\caption{
Same as Figure \ref{fig_lvg_orikl}, but for the OMC3 region.
\label{fig_lvg_omc3}}
\end{figure}

\begin{figure}
\plotone{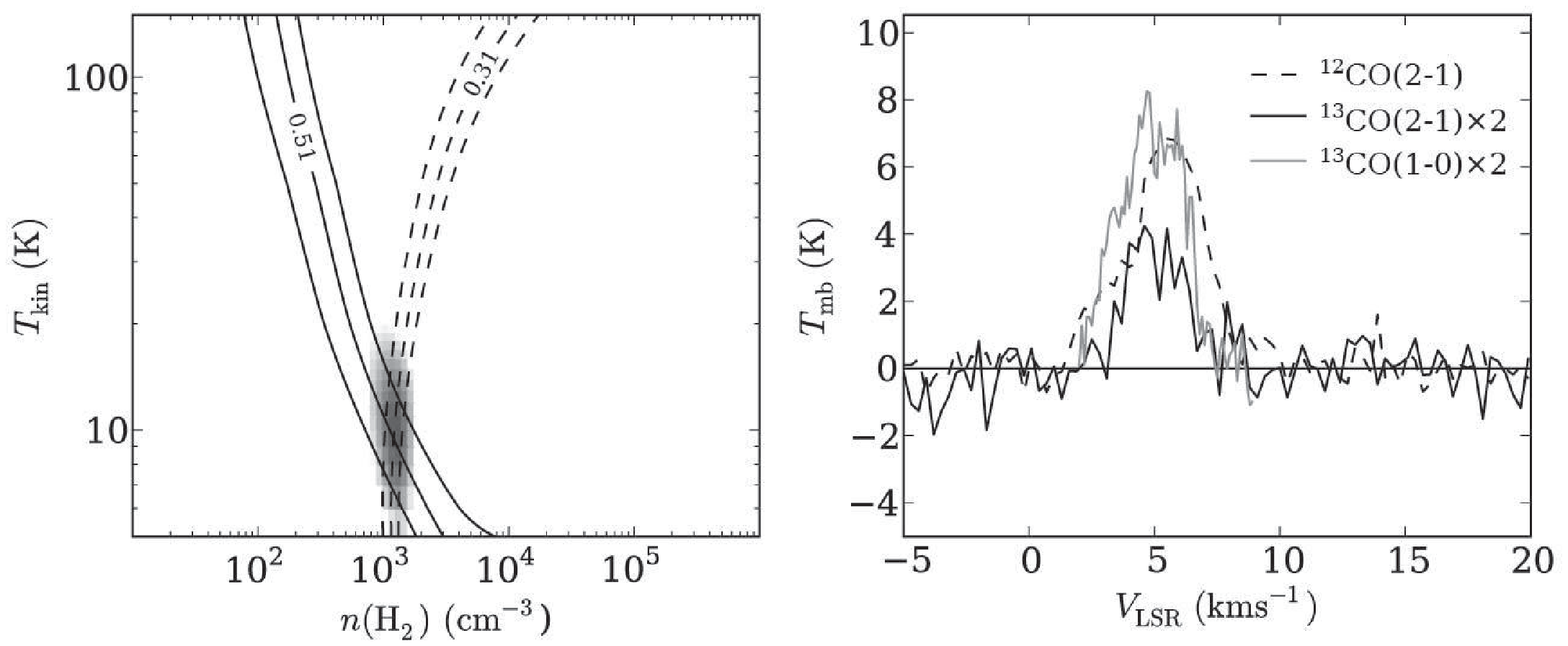}
\caption{
Same as Figure \ref{fig_lvg_orikl}, but for the L1641S region.
\label{fig_lvg_l1641s}}
\end{figure}

First, we chose 7 different points which have different environments for calculating the physical properties with the LVG analyses. 
Orion KL is an example of a region of high temperature (Figure \ref{fig_lvg_orikl}).
This region has a high \Rul \ and low \Rtt.
The analyzed curves are well crossed, and thus the temperature and the density are well determined to be 88 K and 1800 cm$^{-3}$, respectively.
The density is well consistent with the values estimated by \citet{cas90} who determined the density to be a few $10^3$ cm$^{-3}$.
OMC-3 region is an example of a region of high-density and moderate-temperature (Figure \ref{fig_lvg_omc3}).
This region has a high \Rtt \ with moderate \Rul.
The analyzed curves are also well crossed for this region, and the temperature and the density are determined to be 34 K and 2200 cm$^{-3}$, respectively.
L1641S is an example of a region of low-density and low-temperature (Figure \ref{fig_lvg_l1641s}).
This region has the low \Rtt \ with low \Rul.
We determined the temperature and density to be 10 K and 1000 cm$^{-3}$, respectively.
From these analyses, the temperature and density are successfully derived for the different environments.
We also analyzed some other regions.
Results are summarized in Table \ref{table_lvg}.


\subsubsection{Spatial distribution of density and temperature}

\begin{deluxetable}{llrrrrr}
\tabletypesize{\scriptsize}
\tablecaption{
Summary of molecular cloud properties
\label{table_discussion}}
\tablehead{
\colhead{} &
\colhead{} & 
\colhead{$\langle N^{13}_{1-0}(\rm H_2) \rangle$} & 
\colhead{$M^{13}_{1-0}$}  & 
\colhead{$\langle T_{\rm kin} \rangle$} &
\colhead{$\langle n(\rm H_2) \rangle$} & 
\colhead{SFE} \\

\multicolumn{2}{c}{Region / Subregion} & 
\colhead{($10^{20}$ cm$^{-2}$)} & 
\colhead{$(10^3 M_\sun$)} & 
\colhead{(K)} & 
\colhead{(cm$^{-3}$)} &  
\colhead{}
}
\startdata
\multicolumn{2}{l}{The entire Orion region}  & 22.4 &    90 & 28.8 & 1000 & 0.037 \\
\multicolumn{2}{l}{The entire Orion A region}  & 20.8 &    59 & 25.4 & 1000 & 0.045 \\
\nodata & A1 & 16.9 &    18 & 14.9 &  870 & 0.025 \\
\nodata & A2 & 23.2 &    41 & 31.8 & 1100 & 0.054 \\
\multicolumn{2}{l}{The entire Orion B region}  & 26.4 &    30 & 35.8 & 1000 & 0.020 \\
\nodata & B1 & 28.4 &    18 & 44.7 &  990 & 0.018 \\
\nodata & B2 & 24.3 &    12 & 25.5 & 1000 & 0.023 \\
\enddata
\end{deluxetable}

\begin{figure}
\plotone{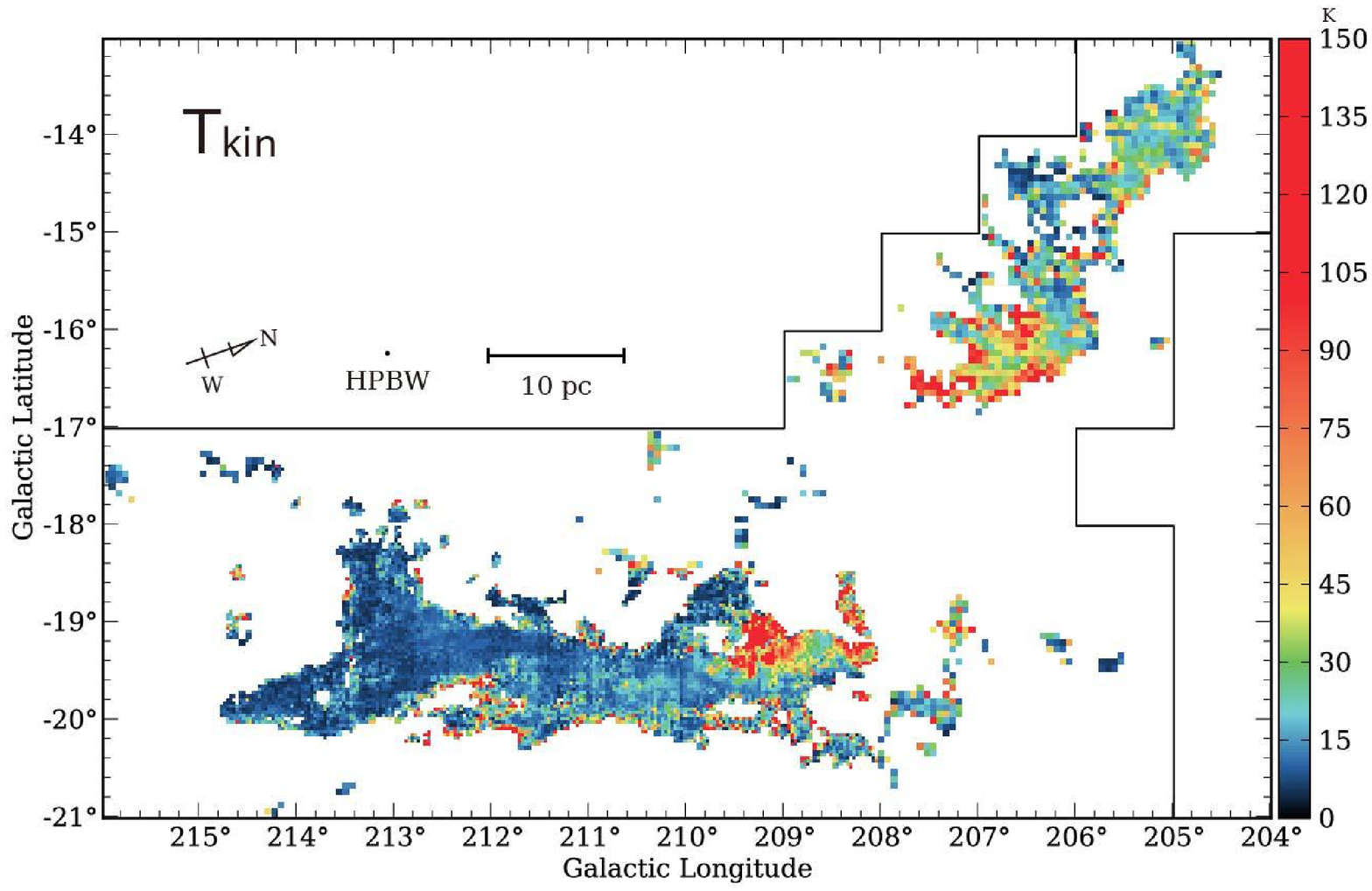}
\caption{
Map of the gas kinetic temperature calculated by the LVG analyses.
The area indicated by the solid line denotes the field observed with the 1.85-m telescope.
(A color version of this figure is available in the online journal.)
\label{fig_lvg_t}}
\end{figure}

\begin{figure}
\plotone{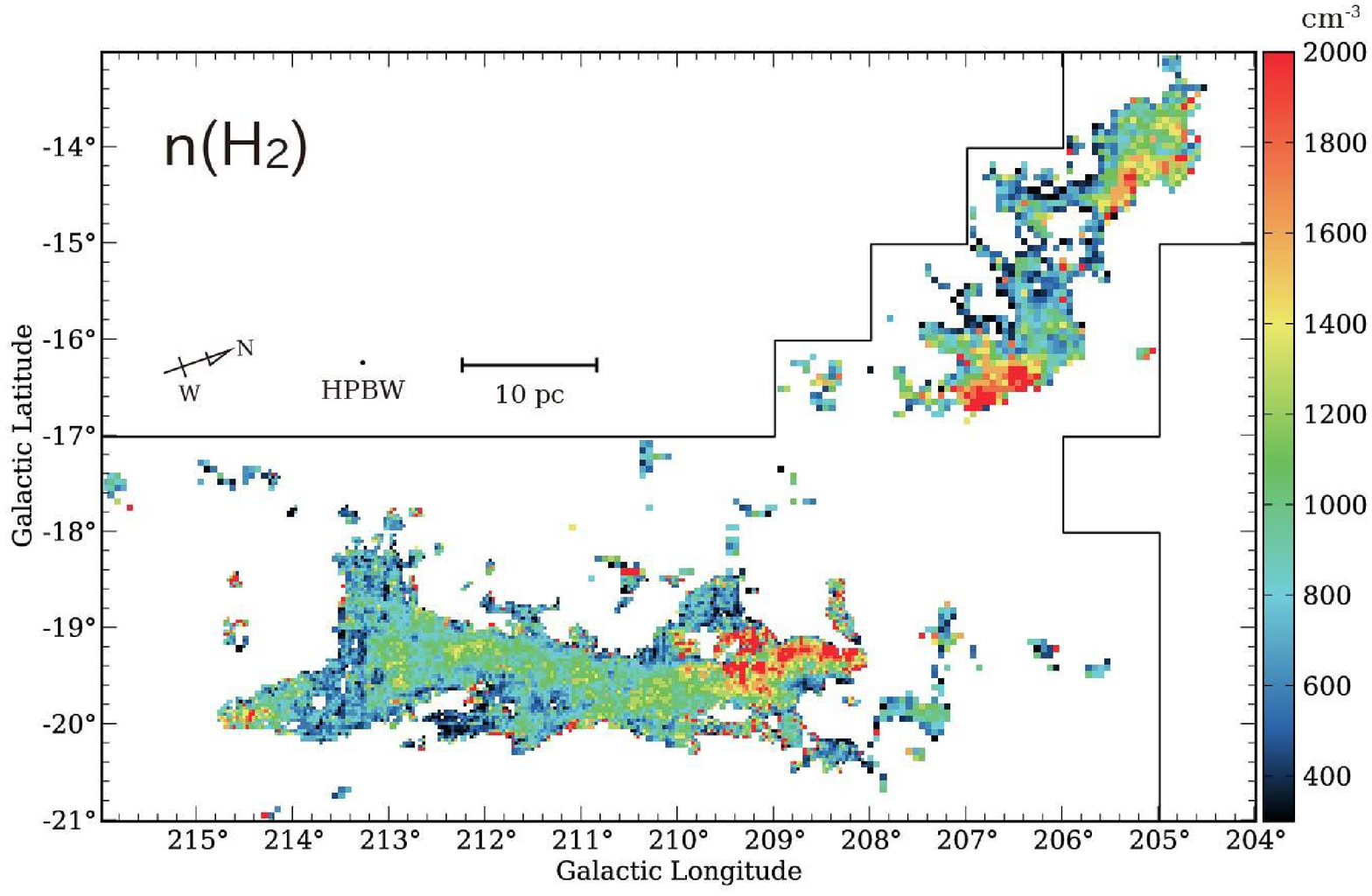}
\caption{
Map of the gas density calculated by the LVG analyses.
The area indicated by the solid line denotes the field observed with the 1.85-m telescope.
(A color version of this figure is available in the online journal.)
\label{fig_lvg_n}}
\end{figure}

As we described in the previous subsection, the temperature and density of the molecular gas can be well determined by the LVG analyses in various environment.
Thus, we apply this method to the whole observed pixels.
Procedures we used are as follows:
(1)we first generated the integrated intensity ratio maps of \Rul \ and \Rtt, and then
(2)we calculated the line intensity for each density and temperature by using LVG analyses assuming uniform sphere structure and constant velocity gradient ($dv/dr$ = 1 km s$^{-1}$ pc$^{-1}$).
(3)We finally compared the observed line ratios and calculated intensity ratios to determine the physical properties of the molecular gas using $\chi^2$ test.
Results of the analyses are shown in Figures \ref{fig_lvg_t} and \ref{fig_lvg_n} for the kinematic temperature and the density of the molecular gas, respectively.

The kinematic temperature is mostly in the range of 20 K to 50 K along the cloud ridge.
The temperature tends to be high in the active star formation sites and decline to the peripheral regions.
We found especially high temperature in some regions. 
One is the east of the Orion KL region near the Trapezium cluster.
This region is considered to interact with the stellar wind and radiation from the Trapezium cluster.
The western part of this region has no significant high temperature structures.
Another region is the southern edge of the Orion B cloud which is located in a front of the OB1b subgroup.
This region seems to be influenced by the radiation of old OB stars.
We also note some other high temperature regions.
One is found in the vicinity of L1641N.
Actually this region has high temperature ($\sim$100 K) but not so large spacial extent as that of Orion KL.
This suggests L1641N is more deeply embedded in the molecular gas.
Another one is the south-west side of the main ridge of the Orion A.
In this region, molecular gas is probably heated by the OB1b or 1c subgroups located southeast to the Orion A molecular cloud.

The densities derived with the analyses show values in the range of 500 to 5000 cm$^{-3}$.
The lowest density we can probe is determined by the critical density of the \tCO($J$=2--1) for excitation.
The highest density we can identify is limited by the thermalization and/or by the high opacity of the line emission toward the dense gas region.
The high density regions ($\sim$2000 cm$^{-3}$) are located in the north of Orion KL for the Orion A and in the south of the Southern cloud for the Orion B.
In the Orion A, the main ridge has a density gradient decreasing toward the outer regions, as pointed out by \citet{sak94}.
We can also find small scale density variations.
For instance, there are local peaks around the L1641N and L1641S regions.



\subsection{Distribution of YSOs}

\begin{figure}
\plotone{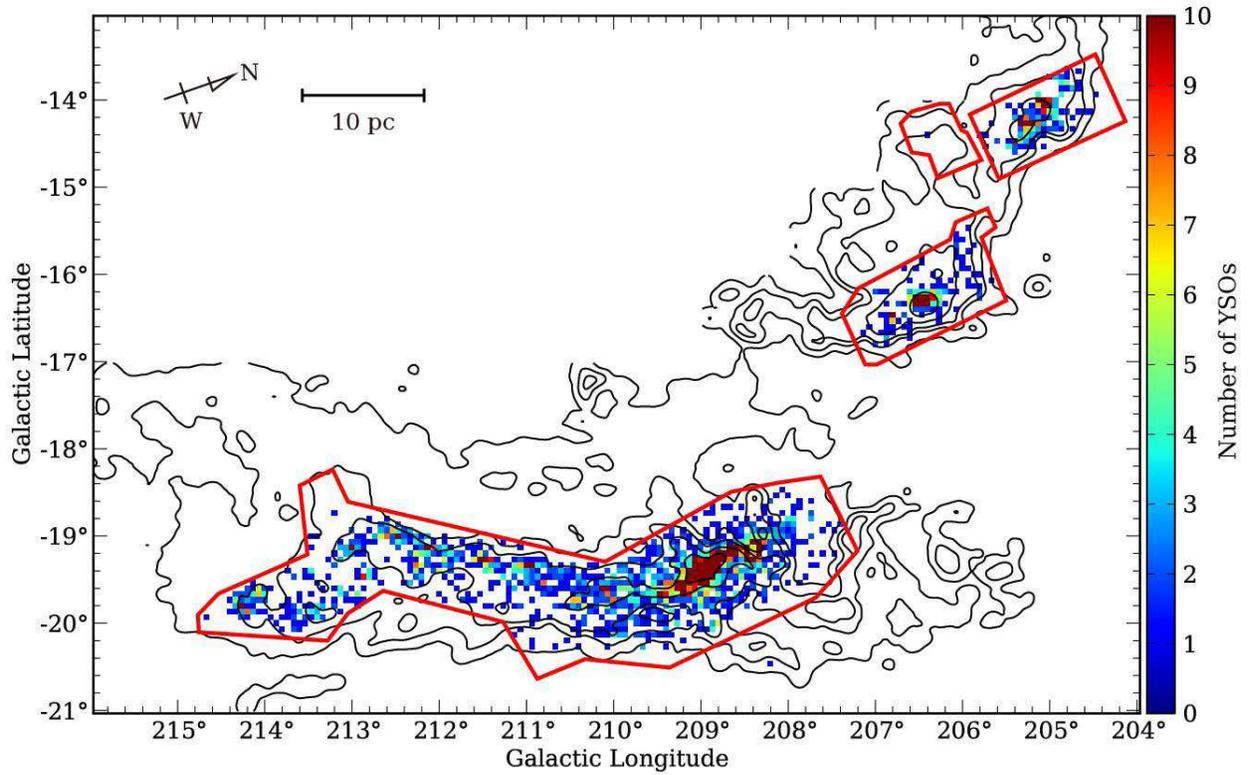}
\caption{
Map of the YSOs surface density\citep{meg12}.
Contours show the integrated intensity of the \CO($J$ = 2--1) smoothed to 10\arcmin (HPBW) for reference.
The contour levels are 2, 10, 20, 50, and 100 K km s$^{-1}$.
The area indicated by the red line denotes the field observed with the Spitzer telescope.
(A color version of this figure is available in the online journal.)
\label{fig_yso_distribution}}
\end{figure}

In this subsection, we compare the derived physical parameters of the gas with the star formation activities.
In the Orion region, a new catalog of Young Stellar Objects (YSOs) was recently complied out of the infrared survey using the Spitzer space telescope \citep{meg12} toward regions with high extinction, which we call Spitzer catalog.
The cataloged YSOs have dusty disk or infalling envelope and then they are considered to be recently formed in the current existing molecular clouds.
The catalog has unveiled the spatial distribution of thousands of YSOs, enabling us to carry out the direct comparison of the Star Formation Efficiency (SFE) with gas temperature and density.

We calculated the surface number density of YSOs, $N_*$, by using the Spitzer catalog at the same grid as our CO dataset (Figure \ref{fig_yso_distribution}).
We used both of the 'disked' and 'protostar' objects in their catalog.
We then derived the distribution of the SFE by the following equation,
\begin{equation}
{\rm SFE} = \frac{M_*}{M_* + M_{\rm cloud}},
\end{equation}
where $M_*$ is the mass of YSOs estimated as $M_* = m_* N_*$ assuming the mean stellar mass $m_* = 0.5 M_\sun$ \citep{eva09}, and $M_{\rm cloud}$ is the mass of the molecular gas.
We use the LTE mass derived from \tCO($J$ = 1--0) line emission for the total molecular gas mass.
We calculated the averaged SFEs for each subregions introduced in \S \ref{sec_mass}.
Results are summarized in Table \ref{table_discussion}.
The subregions in the Orion A have higher SFE than those of the Orion B subregions.



\section{Discussion}


\subsection{Relationship of the cloud physical properties and star forming activity}

\begin{figure}
\plotone{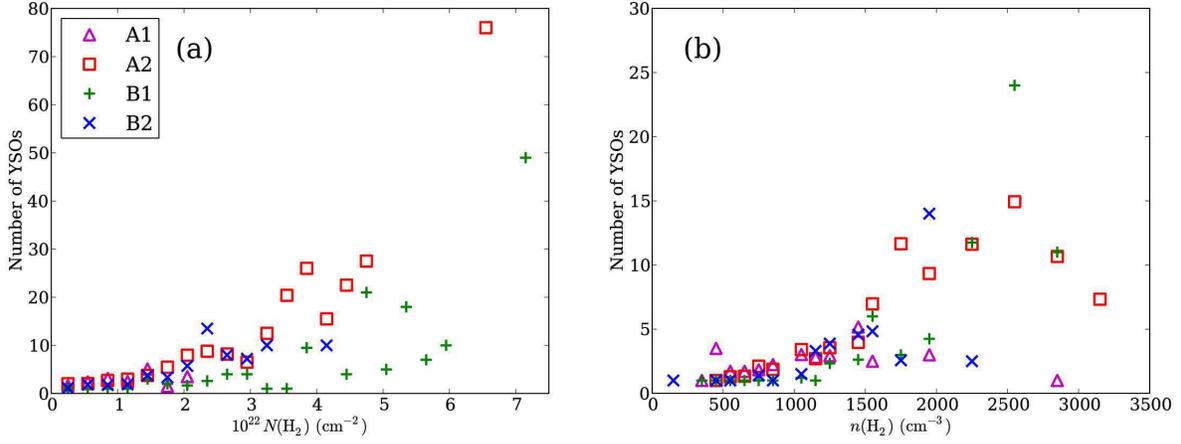}
\caption{
Plot of the average number of YSOs versus (a) the column density and (b) the volume density.
Triangles, squares, plus signs, and crosses denotes the regions A1, A2, B1, and B2, respectively.
(A color version of this figure is available in the online journal.)
\label{fig_discussion_yso}}
\end{figure}

\begin{figure}
\plotone{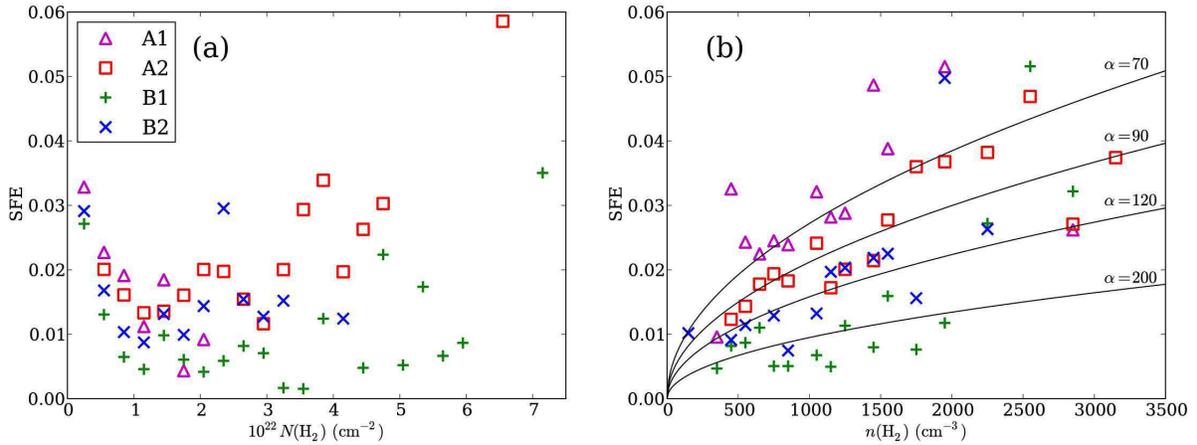}
\caption{
Plot of the average SFE versus (a) the column density and (b) the volume density.
Symbols are same as Figure \ref{fig_discussion_yso}.
The solid lines indicate the relationships of ${\rm SFE} = 0.06 \alpha^{-1} n^{1/2}$ for $\alpha = 70$, 90, 120, and 200, respectively.
(A color version of this figure is available in the online journal.)
\label{fig_discussion_sfe}} 
\end{figure}

In this subsection, we discuss the relationship between the cloud properties and the star formation activities in the Orion molecular cloud.
Figure \ref{fig_yso_distribution} clearly shows that there are more YSOs in regions where the gas column density is higher.
Figure \ref{fig_discussion_yso}a shows that the number of YSOs are positively correlated with the gas column density, although the tendency is unclear in the case of the Orion B2.  
This trend is
also seen in the gas density as shown in Figure \ref{fig_discussion_yso}b, indicating that the density of the gas is a key for the activity of the star formation therein.  
Table  \ref{table_discussion} summarizes the SFEs of the subregions.
It is very striking that the SFE of the Orion A2 subregion is much higher compared with the other subregions.  
However the average column density, temperature, and density of the Orion A2 are not significantly different from the other subregions, implying that the SFE is not necessarily determined by the overall properties of the molecular clouds.  
Figure \ref{fig_discussion_sfe}a shows the relation of SFE with the gas column density, and Figure \ref{fig_discussion_sfe}b with the gas volume density. 
It is obvious that the SFE is well correlated with the gas density, i.e., more stars are formed in denser regions.  
The poorer correlation in Orion B1 may be a result of the gas dispersion due to the active star formation in NGC2024 and the external disturbances as discussed in the next subsection.

The positive correlation between the gas number density and the SFE indicates that the time scale from gas to protostar, $T_{\rm collapse}$, is shorter for denser gas if we assume a steady star formation.
In this case, the total mass of the formed stars $M_*$ is proportional to the total gas mass $M_{\rm cloud}$ and inversely proportional to the time scale of star formation $T_{\rm collapse}$ (i.e., $M_* \propto M_{\rm cloud}/T_{\rm collapse}$).
Therefore, the SFE is inversely proportional to $T_{\rm collapse}$.
If $T_{\rm collapse}$ is described as $\alpha T_{\rm ff}$, where $T_{\rm ff}$ is the free fall time scale, the SFE is proportional to $(\alpha T_{\rm ff})^{-1}$, and then to $\alpha^{-1}n^{1/2}$, where $n$ is the volume density of the gas.
The $\alpha$ depends on the balance among the self-gravity and the other forces, unity for self-gravity dominating case, and then $T_{\rm collapse}$ may depend on the volume density.
The data in Figure \ref{fig_discussion_sfe}b shows that the SFE is roughly explained as ${\rm SFE} \propto n^{1/2}$, although the scatter is large, and the scatter suppose that the dynamics of the gas is different from region to region.

The gas temperature is the highest toward the region around the Orion KL, probably due to the heating by massive stars forming therein.
There is a slight temperature enhancement along the ridge of Orion A, and this may be due to the star formation inside.
We see the enhancement of gas temperature toward NGC2023 in Orion B, although the gas temperature seems not to be well correlated with the star formation activities in the Orion B.

It is to be noted that the Spitzer catalog of \citet{meg12} has not covered the whole extent of the molecular gas.  
A recent study with Akari and WISE cataloges indicates that there are YSOs identified outside the Spitzer area \citep{tot13}.  
We are also interested in the star formation efficiency in somewhat isolated clumps like the EC clumps and the Northern clumps, and this is one of subjects in a subsequent paper.



\subsection{Effect of the surrounding environment}

\begin{figure}
\plotone{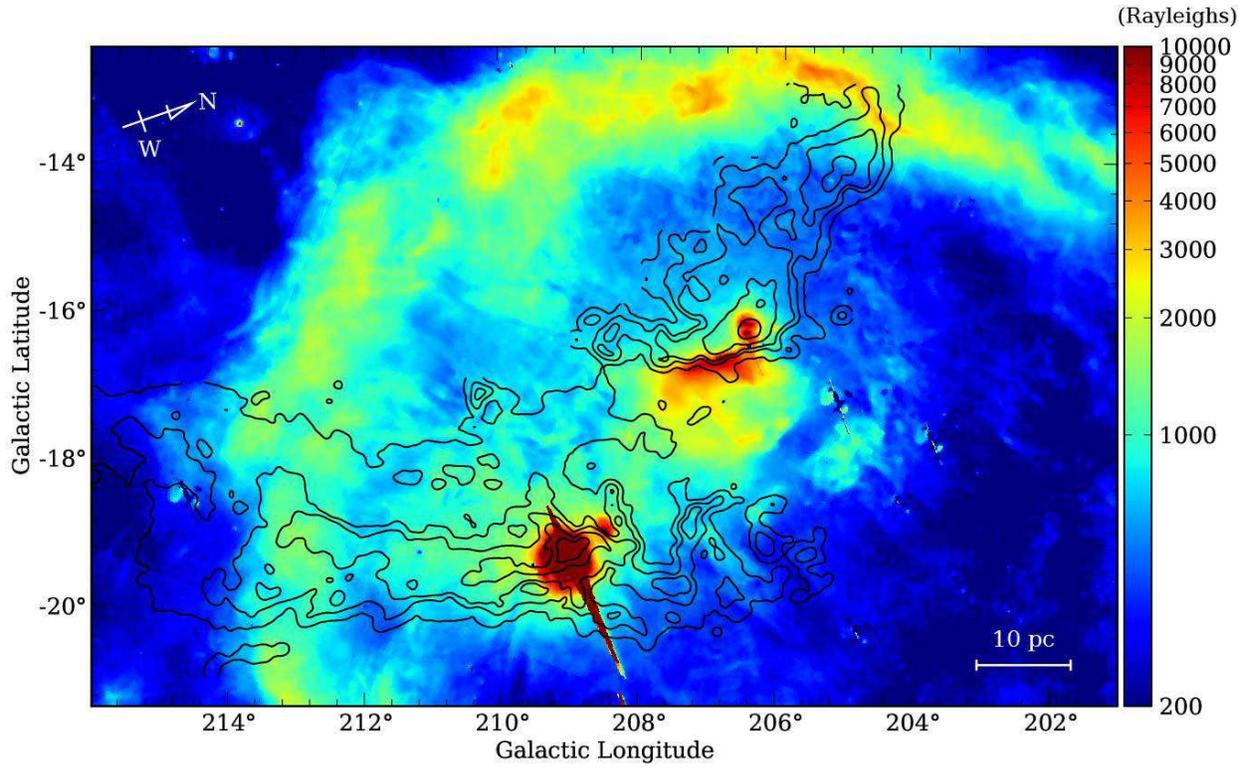}
\caption{
Distribution of the H$\alpha$ intensity \citep{gau01} superposed on the contour of integrated intensity of the \CO($J$ = 2--1) which smoothed to 10\arcmin (HPBW) resolution for reference.
The contour levels are 2, 10, 20, 50, and 100 K km s$^{-1}$.
\label{fig_discussion_halpha}}
\end{figure}

In this subsection, we discuss the effect of the surrounding environment on the physical properties of the molecular clouds.
Figure \ref{fig_discussion_halpha} shows the intensity distribution of \Ha \ compared with the molecular gas distribution.
There are some intense peaks which corresponding to Orion KL, the southern side of the Orion B cloud, and the Bernard loop.  
The Bernard loop is considered to have formed by the interaction with an old supernovae, and the other \HII \ regions are considered to have formed by the Ori OB1 association.
The Bernard loop seems to have no interaction with the molecular cloud as suggested by \citet{sak94}.

The \Ha \ peak toward the Orion KL is clearly due to the current active massive star formation therein.  
The \Ha \ enhancement toward the southern side of the Orion B consists of two parts; one is NGC2024 and the other is along the southern edge of the Orion B cloud. 
The former seems to reflect the ongoing star formation activities.
The latter is ionized by the strong UV radiation from the OB1b subgroups.  
There is a clear gas density and temperature enhancement toward the southern edge of the Orion B1 as shown in Figures \ref{fig_lvg_n} and \ref{fig_lvg_t}, and this fact suggests that the strong stellar wind and UV radiation compress and heat the molecular gas.  
Another important feature of the Orion B1 is that the gas temperature is higher than other subgroups as a whole.  
Especially, the temperature is higher toward the surrounding edge of the Orion B1.  
This implies that the Orion B1 cloud is surrounded by the \HII \ region, heating the outer edge of the molecular gas of the Orion B.



\section{FITS files}

Spectral data of the \CO($J$ = 2--1), \tCO($J$ = 2--1), and \CeO($J$ = 2--1) emission lines are available in FITS format at our website 
URL: \url{http://www.astro.s.osakafu-u.ac.jp/~nishimura/Orion/}.


\section{Summary}

We have observed the $J$ = 2--1 lines of \CO, \tCO, and \CeO \ toward almost the entire extent of the Orion A and B molecular clouds.
By comparing with $J$ = 1--0 data of \CO, \tCO, and \CeO \ observed with the Nagoya 4-m telescopes, we derived the spatial distribution of the physical properties of the molecular clouds and discussed the relation between the cloud physical properties and its surrounding conditions and star formation in the clouds.
The main results are summarized as follows.

\begin{enumerate}

\item
The spatial distributions of each $J$ = 2--1 emission globally resembles that of the corresponding $J$ = 1--0 emission although we observe some differences which reflects the difference in the physical properties.
The general trend is that the distribution of each $J$= 2--1 emission is similar to that of the corresponding $J$ = 1--0 emission toward the region with the high-intensity, although each $J$ = 1--0 line is more widely distributed than that of the corresponding $J$ = 2--1 line toward the region of low-intensity.

\item
The complicated velocity structures are evident in the Orion molecular cloud complex.
Various features are identified in spatial and velocity distributions of these lines.
The Orion A cloud (L1641) includes the main ridge (containing OMC2/3, Orion KL, L1641N, L1641S, and NGC1999), northern clumps, and extended components/clumps.
The Orion B cloud (L1630) includes northern cloud (containing NGC2068, NGC2067), southern cloud(NGC2023, NGC2024), and the 2nd component.

\item
The \CO($J$ = 2--1)/\CO($J$ = 1--0) intensity ratio ($R^{21}_{2-1/1-0}$) is greater than unity in the regions close to the \HII \ region or the cloud boundary facing to the OB association.
This high ratio can be explained if the \CO \ lines are optically thin, and the emitted region is dense to excite to the  $J$ = 2 level and is also warm.
This fact suggests the interaction of the radiation or stellar winds from the massive stars.

\item
We derived the gas mass from the observed line intensities.
We used X-factor of $1.8 \times 10^{20}$ cm$^{-2}$ K$^{-1}$ km$^{-1}$ s \citep{dam01} for optically thick lines of \CO($J$ = 1--0)
and assumed the LTE condition for the optically thin lines of \tCO($J$=1--0), \tCO($J$ = 2--1), \CeO($J$ = 1--0), and \CeO ($J$ = 2--1).
The X-factor masses are similar to that derived from \tCO($J$ = 1--0) intensity
.
The mass derived from $J$= 2--1 is lower than that from $J$= 1--0 by a factor of about 3.
This indicates that the $J$ = 2--1 optically thin lines are sub-thermally excited, and trace denser gas than $J$ = 1--0 lines.

\item 
The spatial distributions of the gas density, $n({\rm H_2})$, and the gas temperature, $T_{\rm kin}$, were derived with the LVG analyses under the assumptions of  the uniform fractional abundance of the CO and the constant $dr/dv$.
The gas temperature is higher in the area around the \HII \ region with $>100$ K.
The gas density is higher ($n({\rm H_2}) > 2000$ cm$^{-3}$)  in the cloud edge facing to the \HII \ region.
These facts suggest the strong stellar wind and UV radiation from the surrounding massive stars are compressing the molecular gas.

\item 
The YSOs surface number density and the SFE are well positively correlated with the gas density.
This fact indicates that the star formation is more effectively taking place in the denser environment.

\end{enumerate}



\acknowledgments

We are grateful to students at Osaka Prefecture University and Tokyo Gakugei University for their help during the observations with the 1.85-m telescope as well as their great efforts in maintaining the telescope.
This work was supported by JSPS KAKENHI Grant Numbers 22244014, 15071205, 22340040, 24244017, 23403001, and 22540250, by JSPS and HAS under the Japan--Hungary Research Cooperative Program, and by the Mitsubishi Foundation.



{\it Facilities:} \facility{1.85-m telescope}, \facility{Nagoya 4-m telescope}




\clearpage


\end{document}